\documentclass[english, aps, prl]{revtex4-1}

\usepackage{amssymb}
\usepackage{amsmath}
\usepackage{color}
\usepackage{ulem}
\usepackage{graphicx}
\usepackage{epstopdf}

\usepackage{bm}
\usepackage{soul}
\usepackage{color}

\def\be{\begin{equation}}
\def\ee{\end{equation}}
\def\bea{\begin{eqnarray}} 
\def\eea{\end{eqnarray}}

\begin{document}

\title{Bi-2212/1T-TaS$_2$ Van der Waals junctions: Interplay of proximity induced high-$T_c$ superconductivity and CDW order}

\author{Ang J. Li, Xiaochen Zhu, G. R. Stewart and Arthur F. Hebard}
\affiliation{Department of Physics, University of Florida, Gainesville, FL 32611, USA}

\begin{abstract}
Understanding the coexistence, competition and/or cooperation between superconductivity and charge density waves (CDWs) in the transition metal dichalcogenides (TMDs) is an elusive goal which, when realized, promises to reveal fundamental information on this important class of materials. Here, we use four-terminal current-voltage measurements to study the Van der Waals interface between freshly exfoliated flakes of the high-$T_{c}$ superconductor, Bi-2212, and the CDW-dominated TMD layered material, 1T-TaS$_{2}$. For highly transparent barriers, there is a pronounced Andreev reflection feature providing evidence for proximity-induced high-$T_{c}$ superconductivity in 1T-TaS$_{2}$ with a surprisingly large energy gap ($\sim 20\thinspace $meV) equal to half that of intrinsic Bi-2212 ($\sim 40\thinspace $meV). Our systematic study using conductance spectroscopy of junctions with different transparencies also reveals the presence of two separate boson modes, each associated with a ``dip-hump" structure. We infer that the proximity-induced high-$T_c$ superconductivity in the 1T-TaS$_2$ is driven by coupling to the metastable metallic phase coexisting within the Mott commensurate CDW (CCDW) phase and associated with a concomitant change of the CCDW order parameter in the interfacial region.   
\end{abstract}

\maketitle
\section*{Introduction}

In the past few decades, the study of interfaces between novel materials including metals, semiconductors, superconductors, topological insulators and layered materials harboring charge density waves (CDWs) has generated the emergence of unexpected phenomena. These discoveries require a new understanding of underlying mechanisms which in turn may well lead to promising new technologies. The proximity effect at the superconducting/normal (S/N) boundary posits a leakage of Cooper pairs into the normal metal accompanied by the appearance of superconductivity at the interface and extending into the metal\cite{deGennes proximity,Deutscher_ASJ reflections,McMillan}. Specifically, detailed study of S/N proximity junctions provides a unique probe of the pairing interaction of both conventional\cite{deGennes proximity} and unconventional\cite{Deutscher_ASJ reflections} superconductors. The reflection of an electron in N as a hole of opposite wave vector (Andreev reflections) and propagation of paired quasiparticles in S gives rise to phenomena including contact-dependent excess conductance, reduced energy gaps and lower transition temperatures on both sides of the interface. Van der Waals (VdW) interfaces advantageously reduce concerns about epitaxial matching across the interface and have already been studied using ultra-smooth cleavable superconductors such as: Bi-2212 in contact with the topological insulators, Bi$_2$Se$_3$ and Bi$_2$Te$_3$\cite{bscco-bi2se3}, 2H-NbSe$_2$ in contact with Bi$_2$Se$_3$\cite{nbse2--bi2se3} and 2H-NbSe$_2$ in contact with graphene\cite{specular andreev}. 
     
In this paper we are motivated by the question of how coexistence, competition or cooperation of superconductivity with various collective electronic states, in particular charge density wave ordered states, can be studied using the proximity effect. This question has been discussed in systems of reduced dimensionality which are prone to electronic instabilities, such as the \textit{coexistence} of CDW order and superconductivity in 2H-NbSe$_{2}$\cite{coexist CDW and SC} in contrast to the \textit{competition} of CDW order and superconductivity in, for example, Yttrium cuprate\cite{compete CDW and SC}. The quasi-two dimensional layered transition-metal dichalcogenides (TMDs), many of which harbor CDW order, have served as model systems for the investigation of the interplay between CDWs and low $T_c$ superconductivity where low-$T_c$ superconductivity can be achieved by the application of pressure\cite{tas2 pressure} to pristine 1T-TaS$_2$ or by partial substitution of Se for S\cite{tas2 atomic origin}, Fe for Ta\cite{tas2 coexist with melted mott} or electric-field gate controlled intercalation of Li\cite{tas2 gate} and is always associated with the more conducting nearly commensurate CDW (NCCDW) phase and even the higher temperature incommensurate CDW (ICCDW) phase. Interestingly, in the pressure experiments on pristine 1T-TaS$_2$, the low-$T_c$ superconductivity ($T_c \approx$ 5\thinspace K) persists until the pressure is high enough to convert the CDW phase to a metal\cite{tas2 pressure}. 

The TMD, 1T-TaS$_{2}$, is a particularly interesting CDW material for such studies because it exhibits a pronounced first order CDW transition with hysteretic resistance transitions in the $180-230$\thinspace K temperature range. It is generally recognized that electron--electron as well as electron--phonon interactions in 1T-TaS$_{2}$\cite{MI1, MI2, MI3, MI4} are responsible for the evolution of a nearly commensurate CDW (NCCDW) to a Mott commensurate CDW (CCDW) phase dominating at low temperatures. Two opposing arguments focus on where and how low $T_c$ superconductivity forms in 1T-TaS$_{2}$. The first argues that the superconductivity is formed within the metallic interdomain spaces separating the CCDW domains where tens of ``star of David'' clusters clump into rough hexagonal domains reproducing the Mott-CCDW phase locally\cite{tas2 pressure, tas2 atomic origin}. The second argues that the superconductivity, characterized by a shallow electron pocket at the Brillouin-zone center, is formed exactly within the clusters of stars\cite{tas2 coexist with melted mott} in the NCCDW phase in real space. We note that low-$T_c$ superconductivity at ambient pressure is only found in non-pristine (i.e., doped) 1T-TaS$_2$ where the NCCDW phase dominates. 

In this study, we find evidence for proximity-induced high-$T_{c}$ superconductivity in the topmost layers of pristine 1T-TaS$_{2}$ at the interface of Van der Waals bonded Bi-2212/1T-TaS$_{2}$ junctions, where pristine Bi-2212 is the high-$T_{c}$ superconductor Bi$_{2}$Sr$_{2}$CaCu$_{2}$O$_{8+\delta}$ with a transition temperature $T_c = 85$~K and intrinsic energy gap $\Delta_0 = 40$~meV. Andreev reflection, marked by excess current and a wide zero-bias conductance peak, is observed at temperatures up to 80~K where the 1T-TaS$_{2}$ is in the Mott-CCDW state. The proximity induced gap $\Delta_a$ in the 1T-TaS$_2$ is found to be a surprisingly large 20\thinspace meV, thereby implying a strong coupling limit ($2\Delta_{sc}/k_{B}T_{c}\sim5.8$) instead of the BCS weak coupling limit of 3.5. In addition to the induced gap in the 1T-TaS$_2$, we also observe a depressed gap compared to the intrinsic gap $\Delta_{0}$  in the vicinity of the interface on the Bi-2212 side. 

Our observation of a transparency-dependent superconducting proximity effect of Bi-2212/1T-TaS$_{2}$ junctions strongly indicates that superconductivity is induced in a metallic phase of 1T-TaS$_{2}$. The transparency is defined as a dimensionless normal conductance, $\sigma_{N}=\frac{1}{1+Z^{2}}$, where the parameter Z is extracted from theoretical fitting of conductance spectroscopy data using the extended BTK model described below. Such a finding is somewhat surprising since, in the absence of intimately contacting Bi-2212, pristine 1T-TaS$_2$ is in an insulating CCDW state, and a proximity effect is only expected to occur in metallic systems. The puzzle here is that at the temperatures where Bi-2212 is superconducting, the stoichiometric pure 1T-TaS$_2$ is in the nonmetallic Mott-CCDW state and a proximity effect is in fact observed thereby implying that the proximity of the Bi-2212 induces changes in the CDW order parameter to achieve a more metallic phase. Said in another way, the mutual interaction of the CDW and superconducting order parameters is such that the proximity-induced gap in the 1T-TaS$_{2}$ can only appear if the CDW order parameter is changed in the interfacial region as is the superconducting gap associated with the Bi-2212. This interpretation is consistent with evidence for a metastable metallic phase, induced by voltage pulses, laser pulses or current excitations, residing within the Mott-CCDW phase\cite{metallic TaS2 1, metallic TaS2 2, metallic TaS2 3, metallic TaS2 4, metallic TaS2 5}. This metastable state is ascribed to the reduction of onsite Coulomb interaction $U$ and increase of Hubbard band width $W$ via phase shifts of the CDW order parameter or interplay between the electron-electron and electron-phonon interactions in the topmost layers\cite{metallic TaS2 1, metallic TaS2 2, metallic TaS2 3, metallic TaS2 4, metallic TaS2 5}.

The superconducting proximity effect, attributed to the leakage of Cooper pairs into a conducting metallic phase of the 1T-TaS$_{2}$, is also confirmed by the good agreement of our \textit{c}-axis conductance spectroscopy $(dI/dV \rm{vs.}$\thinspace$V$) with an extended BTK model for \textit{d}--wave superconductors\cite{BTKhtc2}.
Additionally, our conductance measurements reveal the presence of two dip--hump structures which can be interpreted to reflect an inherent electron--phonon interaction in 1T-TaS$_{2}$ that assists the formation of high-$T_c$ superconductivity in the metastable metallic domains residing within the Mott-CCDW phase. This evidence of electron--phonon interaction assisted high-$T_{c}$ superconductivity within the Mott CCDW phase of 1T-TaS$_{2}$, presents a new paradigm for understanding the correlation of CDW order and high-$T_c$ superconductivity in 1T-TaS$_{2}$.

\section*{Results}
\subsection*{Characteristics of Samples}

Nearly optimally doped crystals of high-$T_{c}$ cuprate Bi$_{2}$Sr$_{2}$Ca$_{1}$Cu$_{2}$O$_{8+\delta}$ (Bi-2212) and stoichiometically pure layered transition metal dichalcoginide (TMD) 1T-TaS$_{2}$ crystals were used. The critical superconducting temperature $T_{c}$ of Bi-2212 (Supplementary Fig.~S.1(a)) and the CCDW-NCCDW phase transition temperatures on cooling/warming processes (Supplementary Fig.~S.1(b)) were verified with AC transport measurements to be at 85\thinspace K and 180\thinspace K/230\thinspace K respectively.

The Bi-2212 and 1T-TaS$_{2}$ crystals were mechanically exfoliated in a dry atmosphere as thin flakes with approximately rectangular shapes and nominal thicknesses of 0.5-2.0\thinspace $\mu$m and 2.0-5.0\thinspace $\mu$m for the Bi-2212 and 1T-TaS$_2$ flakes respectively. The cleaved surfaces were clean and flat, with a mean roughness of 1.55\thinspace $\rm{\AA}$ for Bi-2212 and 1.52\thinspace $\rm{\AA}$ for 1T-TaS$_{2}$ from AFM images as shown in Fig.~\ref{figure1}(c). Two cleaved thin flakes were placed against each other and naturally bonded via Van der Waals forces\cite{van der waals}; high quality normal metal-superconductor (NS) or normal metal-insulator-superconductor (NIS) junctions were then formed. Four terminal tunnel junction configurations with perpendicularly oriented top and bottom electrodes shown in the Fig.~\ref{figure1}(d) schematic have the advantage that contact resistances are eliminated and the active area common to both electrodes (0.2\thinspace $\rm{mm^{2}}$) can be accurately calculated\cite{4wire, 4wire AFH}. Advantageously the \textit{c}-axis of both the bottom (Bi-2212) and top (1T-TaS$_2$) electrodes are perpendicular to the substrate and thus colinear, thereby minimizing currents flowing along the \textit{ab}-plane. Our \textit{c}-axis conductance spectroscopy measurements were performed in a Quantum Design Physical Properties Measurement System (PPMS) at temperatures ranging from 2.5\thinspace K to 120\thinspace K. Using this set up we find the intrinsic Bi-2212 superconducting gap to be in the range 38 to 42\thinspace meV for both Bi-2212/1T-TaS$_{2}$ junctions (see below) and Bi-2212/graphite (Supplementary Fig. S.~2) junctions, in good agreement with previous point-contact tunneling studies on single crystal Bi-2212\cite{Fischer BSCCO gap, Fischer BSCCO gap1}.

\subsection*{Experimental measurements of Bi-2212/1T-TaS$_{2}$ junctions}
\subsubsection*{Highly transparent junctions}
C-axis AC differential conductance spectroscopies and DC  current-voltage ($I-V$) characteristics were measured on Bi-2212/1T-TaS$_{2}$ junctions. To exemplify the superconducting proximity effect at the interface, junctions with high transparency\cite{deGennes proximity, McMillan, highly trans, BTK} or low Z parameter described in the BTK model\cite{BTK} are needed. Under these conditions, Cooper pairs in the superconductor (S) can leak into the normal material (N) side resulting in a spatially dependent superconducting gap $\Delta(x)$  that extends across the interface along the \textit{c}-axis and into the 1T-TaS$_2$. Various gap features can be identified beginning with the intrinsic superconducting gap $\Delta_{0} = 40 $\thinspace mV deep in the superconductor which decreases to a depressed superconducting gap $\Delta_{p} = 28$\thinspace mV at the S/N interface and then decreases to a proximity effect induced superconducting gap $\Delta_{a} = 20$\thinspace mV on the N side of the interface which decreases in magnitude with increasing distance from the interface\cite{deGennes proximity, McMillan}. Moreover, the different proximity region widths within which these gaps exist depend sensitively on the transparency of junctions\cite{deGennes proximity, McMillan, deGennes, Wolf proximity, thickness new method}. In addition, different quasi-particle lifetimes or scattering rates at the vicinity of interface might affect the spectroscopic line shapes\cite{gamasmear}. 

DC current-voltage $I-V$ curves of Junction-1, with BTK parameter Z=0.25, show clear zero-bias excess current relative to the normal current (indicated by the red vertical arrows labeling the induced gap, $\pm \Delta_a$) as shown in Fig.~\ref{figure2}(a) for temperatures from 5\thinspace K to 100\thinspace K. The slope within the excess current range from -20\thinspace meV to 20\thinspace meV at 5\thinspace K, as indicated by the solid gray triangle, is nearly twice the slope within the normal current range indicated by the solid hollow triangle. At temperatures well below the $T_{c}$ ($\sim$85\thinspace K) of Bi-2212, the excess current in the vicinity of zero bias is approximately twice the normal current, and therefore attributable to Andreev reflection at the interface\cite{BTK}, where the electrons injected from N side are reflected as holes with time reversal symmetry tracing the injected electrons' track back into N side. To conserve current across the interface, Cooper pairs flow at the Fermi energy within S.  These Cooper pairs have electron-like (ELQ) and hole-like (HLQ) quasiparticle character above/below the edges of the energy gap and the conversion process only occurs when the incident electrons on the N side lie within the conductance plateau defined over the energy range $\pm \Delta_a$ shown in the normalized differential conductance plot of Fig.~\ref{figure2}(b).  

The two additional nonlinear features marked as blue and magenta vertical arrows in Fig.~\ref{figure2} are attributed respectively to the two superconducting gaps $\Delta_{p}$ and $\Delta_{0}$ mentioned at the beginning of this section. The AC differential conductance spectroscopies $(dI/dV)_{S}$ of Junction-1, normalized by the normal state conductance $(dI/dV)_{N}$ at 100\thinspace K, at various temperatures from 5\thinspace K to 100\thinspace K shown in Fig.~\ref{figure2}(b) confirm the features observed from $I-V$ curves. The proximity effect induced superconducting gap $\Delta_{a}=\pm 20\thinspace \rm{meV}$ at 5\thinspace K on the 1T-TaS$_{2}$ side of the interface delineates the voltage region within which Andreev reflection occurs. Additional gap features, relevant to the nonlinear features observed from $I-V$ curves, are believed to be the density of states (DOS) features corresponding to the depressed superconducting gap $\Delta_{p}$ on the Bi-2212 side at the interface and the intrinsic superconducting gap $\Delta_{0}$ of Bi-2212. For Junction-1, the sizes of the two gaps are $\Delta_{p}=28\thinspace \rm{meV}$ and $\Delta_{0}=40\thinspace \rm{meV}$ respectively at 5\thinspace K. As temperature increases, the zero-bias conductance peak evolves from a flat mesa to a rounded hump while the peak's width and height are depressed up to temperatures near 80\thinspace K. The gaps, $\Delta_{p}$ and $\Delta_{0}$ merge with increasing temperature until all gaps disappear near 80\thinspace K. Similar features are also observed on Junction-2 with BTK parameter Z=0.27, as shown in Fig.~\ref{figure2}(c) inset for DC $I-V$ curves and (d) for AC differential conductance spectroscopies. 

Both Junction-1 and Junction-2 show characteristics of highly transparent junctions, exhibiting  high normal state conductance of 1.85\thinspace $\rm{\mu A/mV}$ and 0.62\thinspace $\rm{\mu A/mV}$ at 5\thinspace K respectively as shown in Fig.~\ref{figure2}.  
As revealed by the conductance spectroscopies of Junction-2 (seen in Fig.~\ref{figure2}(d)), the sizes of the three gaps $\Delta_{0} = 39$\thinspace mV, $\Delta_{p}= 23$\thinspace mV and $\Delta_{a}= 16$\thinspace mV at 5\thinspace K are close to the values of Junction-1. The consistency of the Bi-2212 intrinsic superconducting gap size measured for Junction-1 and Junction-2 with previous \textit{c}-axis point-contact tunneling studies on single crystal Bi-2212\cite{Fischer BSCCO gap, Fischer BSCCO gap1} is satisfying. Any variation of the differential conductance spectroscopies' line-shape for Junction-1 and Junction-2 might be ascribed to the discrepancy of proximity regions' width and scattering rate of quasiparticles at the vicinity of the interface\cite{deGennes proximity, McMillan, deGennes, Wolf proximity, thickness new method, gamasmear}.

Andreev reflection such as observed here for our low-Z junctions (Junction-1 and Junction-2) exists at the normal metal-superconductor (N-S) interface, thereby indicating that the observed Andreev reflection feature on our Bi-2212/1T-TaS$_{2}$ junctions implies at minimum a metallic component on the N side. We conclude that the metastable metallic phase of 1T-TaS$_{2}$ residing in the Mott-CCDW state at low temperatures and revealed by STM studies\cite{metallic TaS2 1,metallic TaS2 2, metallic TaS2 3, metallic TaS2 4, metallic TaS2 5} is the requisite metallic state for proximity coupling to pristine 1T-TaS$_2$, where the metallic phase has a smaller parameter $U/W$ than the Mott-CCDW phase. Importantly, the Andreev enhanced zero-bias peak disappearing at high temperature around 80\thinspace K in Junction-1 and Junction-2 reflects \textit{proximity induced high-$T_{c}$ superconductivity} in the metastable metallic phase residing in the layers of 1T-TaS$_{2}$ which are adjacent to the Bi-2212.

Besides the evidence for a robust superconducting proximity effect existing in our two low-Z Bi-2212/1T-TaS$_{2}$ junctions, we find two dip-hump structures positioned at $\rm{U_{e-ph}}$ and $\rm{U^{*}_{e-ph}}$ for both junctions, which are indicated at their respective peaks (humps) by orange diamond and purple inverted triangles in Figs.~\ref{figure2}(b) and \ref{figure2}(d). As suggested by neutron resonance and ARPES results\cite{neutron resonance diphump, arpes diphump1}, a single dip-hump structure in Bi-2212 probably stems from the electron-boson coupling, or combined electron-boson coupling and pseudogap\cite{arpes diphump2} corresponding to a boson mode energy $\Omega$. 
As the \textit{fingerprint} of a boson mode, the energy at the maximum slope of the dip-hump structure in conductance spectroscopies or the peak/dip in the $dI^{2}/dV^{2}$ spectra at positive/negative bias regions for NIS junctions is positioned at $\rm{E_{p}}=\Delta_{0}+\Omega$\cite{boson mode energy0, boson mode energy1, boson mode energy2, boson mode energy4}. Indicated by orange and purple arrows for the lowest temperatures in Figs.~\ref{figure2}(b) and (c) respectively in the conductance spectroscopies, the determination of experimental values of $\rm{E_{p}}$ and $\rm{E^{*}_{p}}$ from $dI^{2}/dV^{2}$ spectra are described in  Fig.~S.4 of Supplementary information.
Consequently, with the information summarized in Table~\ref{table} at 5\thinspace K, the boson mode energies referred to the two dip-hump structures are $\Omega=\rm{E_{p}}-\Delta_{0}=48\thinspace \rm{meV}$ and $\Omega^{*}=\rm{E^{*}_{p}}-\Delta_{0}=24\thinspace \rm{meV}$ for Junction-1 with $\Delta_{0}=40\thinspace \rm{meV}$; $\Omega=40\thinspace \rm{meV}$ and $\Omega^{*}=17\thinspace \rm{meV}$ for Junction-2 with $\Delta_{0}=39\thinspace \rm{meV}$. For intrinsic Bi-2212, the boson mode energy found in STM spectra is around 52$\pm$8\thinspace meV\cite{stm mean boson mode energy}, leading us to conclude that the hump peaked at $\rm{U_{e-ph}}$ is from intrinsic Bi-2212 accompanied by a boson mode energy $\Omega$, while the hump peaked at $\rm{U^{*}_{e-ph}}$ is a heretofore unseen hump due to another boson mode with lower energy $\Omega^{*}$. As temperature increases, the two humps merge to a single hump near 50-60\thinspace K for both junctions, a temperature somewhat more than the temperature near 40\thinspace K where $\Delta_{0}$ and $\Delta_{p}$ merge as shown in the differential conductance spectroscopies in Figs.~\ref{figure2}(b) and (d).

\subsubsection*{Two boson modes}
 
Our Bi-2212/1T-TaS$_{2}$ Junction-3, with slightly lower transparency (BTK parameter Z=0.5), offers more information about the dip-hump structures observed in Junction-1 and Junction-2. As shown in Fig.~\ref{figure3}(a), the DC $I-V$ curves at various temperatures (3\thinspace K to 30\thinspace K) well below Bi-2212 $T_{c}$ show several nonlinear features. Besides the nonlinear feature of the Bi-2212 intrinsic superconducting gap $\Delta_{0}$ indicated by the magenta vertical arrow, the other two nonlinear features taking place above $\Delta_{0}$, as indicated by the purple and orange vertical arrows, are also clearly observed. Junction-3 with lower normal conductance around 0.21\thinspace $\rm{\mu A/mV}$ at 5\thinspace K than Junction-1 and Junction-2, however, demonstrates the normal conductance varies with temperatures indicating the 1T-TaS$_{2}$ for Junction-3  might behave more like a Mott-CCDW material rather than the metastable metallic material in the more transparent junctions. The temperature dependence of normal conductance in Junction-3, referring to a less metallic NS junction, probably originates from the carrier delocalization in the Mott-insulating state in 1T-TaS$_{2}$ with increasing temperature\cite{MI1, MI2, MI3, MI4}.

The normalized AC differential conductance spectroscopies $(dI/dV)_{S}$/$(dI/dV)_{N}$ of Junction-3, obtained using the method described in the Methods section, at various temperatures from 3\thinspace K to 100\thinspace K are shown in Fig.~\ref{figure3}(b). A strongly suppressed amplitude of the zero-bias peak for Junction-3 relative to Junction-1 and Junction-2 is observed. Such strong suppression suggests a stronger scattering rate or shorter quasiparticle lifetimes in the proximity-effect-induced superconducting region on the 1T-TaS$_{2}$ side at interface. The width of the zero-bias peak decreases with increasing temperature, and is totally suppressed near 40\thinspace K. Meanwhile, the feature representing the depressed superconducting gap $\Delta_{p}$ is clearly not observed,
as seen in Fig.~\ref{figure3}(b),
which we ascribe to the closer proximity between $\Delta_{p}$ and $\Delta_a$ and temperature broadening. For Junction-3, the measured superconducting gap $\Delta_{0}$ around 38\thinspace meV at 5\thinspace K is slightly smaller but still consistent with the highly transparent Junction-1 and Junction-2. Notably, the width of zero-bias peak is around 21\thinspace meV, which is not strongly depressed relative to Junction-1 and Junction-2.

Sharper and clearer signatures of two dip-hump structures are observed in Junction-3. The conductance spectroscopy indicates the hump (peak marked by the orange diamond) due to the boson mode energy $\Omega$ has a broader width than the hump (peak marked by the purple inverted triangle) due to the boson mode energy $\Omega^{*}$. The two humps of Junction-3 at 5\thinspace K which are peaked at the energy $\rm{U_{e-ph}}=93\thinspace \rm{meV}$ and $\rm{U^{*}_{e-ph}}=66\thinspace \rm{meV}$, as well as the maximum slope positions of dip-hump structures $\rm{E_{p}}=89\thinspace \rm{meV}$ and $\rm{E^{*}_{p}}=62\thinspace \rm{meV}$ are listed in Table~\ref{table}, which correspond to the boson mode energies $\Omega = \rm{E_p}-\Delta_0 =89-38=51$\thinspace meV and $\Omega^{*} = \rm{E^*_p} - \Delta_0 = 62-38=24$\thinspace meV. More details about the temperature evolution (at temperatures below Bi-2212 $T_{c}$) of the Bi-2212 superconducting gap $\Delta_{0}$ and two humps are presented in Fig.~\ref{figure3}. Similar to the highly transparent Junction 1 and 2, the two humps merge toward each other with increasing temperature, and almost merge into a single hump at an energy position of $\rm{U_{e-ph}}=\rm{U^{*}_{e-ph}}=39\thinspace \rm{meV}$ at around 60\thinspace K. The zero-bias conductance peak is totally suppressed at a lower temperature around 40\thinspace K instead of 80\thinspace K as in Junction 1 and 2; however, still indicating proximity-induced high-$T_{c}$ superconductivity in 1T-TaS$_{2}$.

\subsection*{Theoretical modeling and results}

The superconducting proximity effect distinguishes two major regimes at the interface of NS or NIS junctions\cite{deGennes proximity, McMillan}. The NS interface for electrons with energy $E<\Delta_{a}$ injected from the N side is the boundary between the induced superconducting region and the normal material on the N side, whereas electrons within energy $\Delta_{a}<E<\Delta_{p}$ and $\Delta_p <E <\Delta_{0}$ will experience the NS interface from the S side. Two schemes of a proximity junction with thicknesses of the proximity regions on the S and N sides respectively marked as $d_{S}$ and $d_{N}$ are depicted in Fig.~\ref{figure4}(a). Schemes Nos.1 and 2 correspond respectively to the cases $\Delta_{0}>\Delta_{p}>\Delta_{a}$ and $\Delta_{0}>\Delta_{p}\sim\Delta_{a}$. 

The \textit{c}-axis conductance characteristics of various junctions, including Bi-2212/1T-TaS$_{2}$ and Bi-2212/graphite junctions, are calculated based on an extended BTK model of the tunneling spectrum for anisotropic superconductors\cite{ BTKhtc2}. For Bi-2212, with $d$-wave symmetry of the superconducting gap\cite{ bsccodwave2, bsccodwave3, bsccodwave4}, the electron-like quasiparticle (ELQ``+") and hole-like quasiparticle (HLQ``--") experience the same pairing potential, $|\Delta_{+}|=|\Delta_{-}|=\Delta_{0}\rm{cos}(2\alpha)$, where the angle $\alpha$ in the plane perpendicular to the \textit{c}-axis is a measure of the gap lobe's orientation and global phase $\phi_{+}=\phi_{-}=0$ along the \textit{c}-axis. Then, the \textit{c}-axis normalized differential conductance $(dI/dV)_{S}$/$(dI/dV)_{N}$ is expressed as

\renewcommand{\theequation}{\arabic{equation}}
\be
\frac{(dI/dV)_{S}}{(dI/dV)_{N}}(V)=\int_{-\infty}^{+\infty}\frac{\partial f_{0}(E-eV)}{\partial (eV)}\sigma_{T}(E)dE~,
\label{didvext}
\ee
where $f_{0}(E)$ is the Fermi-Dirac distribution at temperature $T$. The dimensionless tunneling conductance at an energy $E$ away from $E_{F}$ is described as

\be
\sigma_{T}(E)=\frac{\int_{\Omega}[1+R_{eh}^{2}(E)-R_{ee}^{2}(E)]\sigma_{N}\rm{cos}\theta d\Omega}{\int_{\Omega}\sigma_{N}\rm{cos}\theta d\Omega}~.
\label{tunneling sigma}
\ee

Equation~(\ref{tunneling sigma}) corresponds to a semi-spherical solid angle integration over the Fermi surface of $d$-wave superconductors. In the case of no mismatch of the Fermi level across the interface, the dimensionless normal conductance or transparency of the interface is $\sigma_{N}=\frac{1}{1+Z^{2}}$ as mentioned earlier. The rate of Andreev reflection (AR) $R_{eh}(E)$ and ordinary reflection (OR) $R_{ee}(E)$, based on the extended BTK model\cite{BTKhtc2}, are expressed in equation (S.5) in Supplementary information. Additionally, by simply adding an imaginary energy term $-i\Gamma$ (as the quasiparticle lifetime parameter) to $E\rightarrow E-i\Gamma$ and $\Omega_{\pm}\rightarrow \sqrt{(E-i\Gamma)^{2}-\Delta_{0}^{2}}$, the smearing effect on tunneling spectroscopy due to quasiparticle life time $\tau_{R}$ or scattering rate $1/\tau_{R}$ is quantitatively described\cite{gamasmear, bscco lifetime}.

The calculated conductance spectroscopies for various junctions show good agreement with measurement results, as shown in Fig.~\ref{figure4}(b) for various junctions at 5\thinspace K, as well as the temperature-dependent conductance spectroscopies below Bi-2212 $T_{c}$ for Bi-2212/1T-TaS$_{2}$ and Bi-2212/graphite junctions seen in Figs.~S.~3(a)-(e). Fig.~\ref{figure4}(c) also compares the temperature dependent gaps ($\Delta_{0}$, $\Delta_{p}$ and $\Delta_{a}$) with theoretical calculations for various Bi-2212/1T-TaS$_{2}$ junctions. These calculations were only carried out up to 40\thinspace K because of high temperature smearing effects. The parameters used in calculation of the conductance characteristics at 5\thinspace K for various junctions are listed in Table~\ref{table}. More details on theoretical modeling are described in Supplementary information (Section: Supplementary Discussion--Theoretical Modeling).

\begin{table}[t]
\centering
\resizebox{\textwidth}{!}{\begin{tabular}{c c c c c c c c c c c c} 
\hline\hline
& Z & $\tilde{\sigma}_{N}$ & $\Delta^{exp}_{0}$/$\Delta^{theory}_{0}$ &  $\Delta^{exp}_{p}$/$\Delta^{theory}_{p}$ & $\Delta^{exp}_{a}$/$\Delta^{theory}_{a}$ & $\Gamma_{0}$ & $\Gamma_{p}$ & $\Gamma_{a}$ & $\rm{U_{e-ph}}$/$\rm{E_{p}}$ & $\rm{U^{*}_{e-ph}}$/$\rm{E^{*}_{p}}$ & $\frac{d_S}{\xi_0}/\frac{d_{vdW}}{\xi_0}/\frac{d_N}{\xi_0}$ \\
& & $\rm{\mu A/mV}$ & $\rm{meV}$ & $\rm{meV}$ & $\rm{meV}$ &\rm{meV} &\rm{meV} & \rm{meV}& $\rm{meV}$ & $\rm{meV}$ &\\
\hline
Junction-1  & 0.25 & 1.85 & 40/41.0 & 28/26.8 & 20/19.1 & 0.8 & 2.7 & 0.1 & 95/88 & 65/64 & 5.5/2/10\\
Junction-2 & 0.27 & 0.62 & 39/39.6 & 23/22.8 & 16/16.5 & 0.8 & 0.2 & 0.1 & 83/79 & 63/56& 5/1.9/8\\
Junction-3 & 0.5 & 0.21 & 38/42.3 & 21/18.5 & 21/17.3 & 0.6 & 9.1 & 5.2 & 93/89 & 66/62&4.4/1.6/2\\
Junction-4 & 0.8 & ---&39/39.2 & 23/23.3 & 23/22.2 & 2.1 & 2.3 & 2.2 & --- & --- & 4/1.2/0.5\\
Junction-5 & 2.5 & ---&41/40.3 & --- & --- & 8.1 & --- & --- & --- & --- &---\\
Bi-2212/graphite & 1 & 55.62 & 42/40.5 &---&---& 6.1 &---&---&---&---&---\\
\hline
\hline
\end{tabular}}
\caption{Parameters used in calculation (marked as ``\textit{theory}'') or from measurements (marked as ``\textit{exp}'') of the conductance spectroscopies at 5\thinspace K for various junctions. Z: BTK parameter; 
$\tilde{\sigma}_{N}$: normal conductance from measurement; $\Delta_{0}$: Bi-2212 intrinsic superconducting gap; $\Delta_{p}$: suppressed superconducting gap on S side; $\Delta_{a}$: proximity induced superconducting gap on N side; $\Gamma_{0}$: quasiparticle lifetime parameter on S side; $\Gamma_{p}$: quasiparticle lifetime parameter at interface on S side; $\Gamma_{a}$: quasiparticle lifetime parameter at interface on N side; $\rm{U_{e-ph}}$/$\rm{U^{*}_{e-ph}}$: position of higher/lower energy scaled hump on conductance spectroscopy. $\rm{E_{p}}$/$\rm{E^{*}_{p}}$: position of higher/lower energy scaled peak/dip in positive/negative voltage regions of $d^{2}I/dV^{2}$. $d_S/d_{vdW}/d_N$: thickness of proximity region on S side, Van der Waals stacking distance and thickness of  proximity region on N side. } 
\label{table} 
\end{table}

\section*{Discussion}

The normalized differential conductance plots, as shown in Fig.~\ref{figure5}(a), at temperatures from 5\thinspace K to 30\thinspace K for Bi-2212/1T-TaS$_{2}$ Junction-4 with BTK parameter Z=0.8 exhibit the proximity effect described by Scheme No.2 in Fig.~\ref{figure4}(a). For this case there is no discontinuity in the gap at the interface (i.e., $\Delta_p = \Delta_a$). For even lower transparency interfaces, such as Junction-5 with BTK parameter Z=2.5 shown in Fig.~\ref{figure5}(b), any remnants of a superconducting proximity effect with markers at $\Delta_p$ and $\Delta_a$ have disappeared. To briefly summarize, for relatively high transparent Bi-2212/1T-TaS$_{2}$ Junction 1--4, the superconducting proximity effect is clearly experimentally observed (seen in Fig.~\ref{figure2}, Fig.~\ref{figure3} and Fig.~\ref{figure5}(a)) via \textit{c}-axis conductance spectroscopies. The temperature dependence of the gaps $\Delta_{0}$, $\Delta_{p}$ and $\Delta_{a}$ for Junction 1--4, displayed in the four panels of Fig.\ref{figure4}(c), show a merging of $\Delta_{p}$ and $\Delta_{a}$ with decreasing transparency (or increasing Z) of the interface and increasing temperature.

From the linear DC $I-V$ characteristics (seen in Fig.~\ref{figure2}(a) and Figs. S.~5(a)-(b)), the normalized zero-bias conductance (NZBC) for Junction-1 with Z=0.25 divided by the conductance at 100\thinspace K is calculated and shown in Fig.~\ref{figure6}(a) as a function of temperature for the cooling cycle. At high temperatures above $T_c$ where the Bi-2212 is in the normal state, hysteresis dominates in the temperature range of 180\thinspace K to 230\thinspace K as it does in the four-terminal temperature-dependent resistance of a pristine 1T-TaS$_2$ flake shown in Supplementary Fig. S.~1(b).

  For additional insight, we use a back-to-back structured Ag/1T-TaS$_2$/Ag trilayer junction to detect the perpendicular transport characteristic of a metal/1T-TaS$_2$ junction. The linear $I-V$ characteristics are shown in Fig S.~5(c)-(d). As shown in Fig.~\ref{figure6}(b), the temperature-dependent NZBC of Ag/1T-TaS$_2$/Ag reveals the signature of the CDW transition in 1T-TaS$_2$ by a strong suppression of the NZBC when 1T-TaS$_2$ transits from a metallic NCCDW phase to a Mott-CCDW phase. In this case the metal/1T-TaS$_2$ junction is expected to include a metal/semiconductor barrier owing to the opened Mott gap of 1T-TaS$_2$ at low temperatures. Thus, the observed hysteresis in Junction-1 suggests the existence of CDW transitions when Bi-2212 is intimately contacted to 1T-TaS$_2$ either as a superconductor or a normal metal. In both cases the 1T-TaS$_2$ has clearly converted to a Mott-CCDW phase at low temperatures. Moreover, the NZBC of Junction-1 in panel (a) has increased with decreasing temperature up to almost 2 times that of the referenced 100\thinspace K conductance due to the Andreev reflection contributions at base temperature 5\thinspace K. Accordingly, the superconducting proximity effect and the CDW transition are simultaneously present in the same sample. We also note an obviously suppressed amplitude of the hysteresis window for Junction-1 compared to the Ag/1T-TaS$_2$/Ag junction. Such a signature of suppression reveals the more metallic nature of 1T-TaS$_2$ when in contact with Bi-2212.

Consequently, the superconducting proximity effect observed in our highly transparent Bi-2212/1T-TaS$_{2}$ junctions strongly suggests that high-$T_{c}$ superconductivity forms within the metastable metallic phase with a smaller parameter $U/W$, residing in the Mott-CCDW phases located in the topmost layers of 1T-TaS$_{2}$. The different heights and widths of the zero bias peak on the conductance characteristics for various Bi-2212/1T-TaS$_{2}$ junctions with different transparencies might be ascribed to the Andreev reflection corresponding to different configurations of Mott-CCDW phase and metallic phase in 1T-TaS$_{2}$ (See Supplementary information for more discussion). The curved and depressed zero-bias conductance peak (seen in Fig.\ref{figure2}(b)\&(d) and Fig.\ref{figure3}(b)) corresponds to the smearing effect with increasing quasiparticle scattering rate\cite{gamasmear, bscco lifetime}, Accordingly, it is no surprise that the quasiparticle lifetime parameters  $\Gamma_a$ at the interface on the N side (see Table~\ref{table}) are lower by more than a factor of ten for the high transparency Junction 1 and 2 than they are for the lower transparency Junction 3 and 4. 

We rule out alternative interpretations with the following arguments: Firstly, the Bi-2212 intrinsic superconducting gap $\Delta_{0}$ measured from various Bi-2212/1T-TaS$_{2}$ and Bi-2212/graphite junctions shows consistency of the results at temperatures well below Bi-2212 $T_{c}$, with the BCS gap ratio $2\Delta_{sc}/k_{B}T_{c}$ to be around 10.4--11.5 (seen in Fig.\ref{figure5}(c)) in good agreement with previous works on intrinsic Bi-2212\cite{Fischer BSCCO gap, Fischer BSCCO gap1}. Moreover, less dependence of the proximity induced gap $\Delta_a$ on temperature (seen in Fig.\ref{figure4}(c)) rules out a contribution to the differential conductance from \textit{Andreev bound states caused by planar geometry or surface roughness}\cite{CRH1, andreev bound state1,rough surface andreev bound1}. (Here, we claim the larger deviation of measured superconducting gaps among various junctions at higher temperatures close to the $T_{c}$ of Bi-2212 is probably due to distortion of gap features on conductance spectroscopies by stronger scattering and temperature smearing effects, which could be self-consistently verified by the smaller deviations of gap parameters used in theoretical results shown in Fig.~S~3(f)). Secondly, though there is a strong suppression of amplitude for the zero bias conductance peak for various junctions (almost two times the normal conductance in Junction-1 and Junction-2, whereas 30\% larger than normal conductance in Junction-3), less temperature dependence of the wide induced gap $\Delta_{a}$ (41\% to 55\%$\Delta_{0}$ on 1T-TaS$_{2}$ side as shown in Fig.\ref{figure4}(c)) rules out the feature of the \textit{complex Andreev reflection due to the phase conjugation of electrons and holes} predicted for superconductor--semiconductor interfaces\cite{phase conjugation}. Thirdly, there is no dominant feature of periodic conductance peaks observed in any of our Bi-2212/1T-TaS$_{2}$ junctions, which rules out the possibility that our observed proximity feature is related to \textit{McMillan--Rowell oscillation} observed in some normal metal/cuprate junctions\cite{Wolf proximity, McMillan-Rowell, McMillan-Rowell1, McMillan-Rowell2,McMillan-Rowell3}. Lastly, with a very short \textit{c}-axis coherence length ($\sim$1\AA \cite{caxis coherence length bscco1}) in intrinsic Bi-2212, the absence of evidence for Josephson junctions intrinsically formed by superconducting CuO$_{2}$ and non-superconducting Bi--O and Sr--O layers\cite{intrinsic Josephson1, intrinsic Josephson2} observed at temperatures well below $T_{c}$ rules out the explanation of the \textit{superconducting proximity effect masquerading as interlayer tunneling} within the Bi-2212.

In Section IIB-2 we discussed the presence of two dip-hump structures, $\rm{U_{e-ph}}$  and $\rm{U^*_{e-ph}}$ which when associated with peak/dip features in the $dI^{2}/dV^{2}$ spectra at positive/negative bias regions defined respectively the energies $\rm{E_{p}}$ and $\rm{E^{*}_{p}}$. We then referenced these features to the intrinsic Bi-2212 gap edge $\Delta_0$ using the relations $\Omega = \rm{E_p}-\Delta_0$ and $\Omega^{*} = \rm{E^*_p} - \Delta_0$ and find that the temperature-dependent boson mode energy $\Omega$ is consistent with the STM spectrum on intrinsic Bi-2212\cite{stm mean boson mode energy}. This picture is incomplete however without comparing the energy of the second feature referenced to the intrinsic gap, $\Omega^{*} = \rm{E^*_p} - \Delta_0$ and alternatively referenced to the reduced gap at the interface, $\Delta_p$, using the relation $\Omega_{p} = \rm{E^*_p} - \Delta_{p}$, revealing a slightly larger energy scale than $\Omega^*$ both of which are derived from $\rm{E^*_p}$. For Junction 1-3 these three boson modes, $\Omega$, $\Omega^{*}$ and $\Omega_p$ are plotted in Fig.~\ref{figure5}(d) as a function of temperature.

Here, we provisionally assume the two dip--hump structures in conductance spectroscopies, observed in Junction-1--3, are due to self energy effects related to electron-boson interaction where phonons serve as the relevant bosons. We then infer that the ``glue'' assisting the strong-coupled pairing of electrons responsible for the high-$T_{c}$ superconductivity in 1T-TaS$_{2}$ arises from phonons. The key point, here, is whether the additional hump peaked at $\rm{U^{*}_{e-ph}}$ is the hump corresponding to the suppressed superconducting gap $\Delta_{p}$ because of a boson mode energy $\Omega_{p}$ representing the difference between the junction interface and bulk Bi-2212, or a hump related to intrinsic superconducting gap $\Delta_{0}$ due to a boson mode energy $\Omega^{*}$ incorporating the effect of interplay between electron-electron and electron-phonon interactions in 1T-TaS$_{2}$ on the density of states. For all junctions, the boson mode energy $\Omega$ at low temperatures is within the range of 40\thinspace meV to 60\thinspace meV (as indicated by the gray background) as well as being less temperature dependent. The results are consistent with the STM spectrum on intrinsic Bi-2212\cite{stm mean boson mode energy}. The yellow background demarcates the possible reduction ($\sim$4\thinspace meV\cite{stm mean boson mode energy}) due to the substitution of $^{16}$O by $^{18}$O in Bi-2212 crystals. Slightly lower energy $\Omega_{p}$ relative to $\Omega$ is observed in all Junction 1--3. On the other hand, all junctions reveal a nearly temperature-independent boson mode energy $\Omega^{*}$ around 25\thinspace meV indicated by the horizontal black dashed line (except Junction-2 over limited temperatures). The magnitude of $\Omega^{*}$ is in good agreement with the energy scale of an observed 25\thinspace meV infrared optical phonon\cite{Tanner TaS2} corresponding to a CDW near the same energy\cite{interplay of e-e and e-ph TaS2} that provides insight into the disorder--induced quasimetallic phase of 1T-TaS$_{2}$ residing in the Mott-CCDW phase. Additionally, the observed evolution of broader (Junction-1--2 as seen in Fig.~\ref{figure2}) to sharper (Junction-3 as seen in Fig.~\ref{figure3}) line-shape of the hump peaked at $\rm{U^{*}_{e-ph}}$ is consistent with current evidence and explanations of the transition from a Mott-insulating to a metallic phase in 1T-TaS$_{2}$ due to reduced Coulomb interaction $U$ and broadened band width $W$ of lower Hubbard band\cite{metallic TaS2 1,metallic TaS2 2,metallic TaS2 3,metallic TaS2 4,metallic TaS2 5,interplay of e-e and e-ph TaS2}. 

In conclusion, we have used differential conductance spectroscopy of Bi-2212/1T-TaS$_2$ junctions with varying transparencies to find high-$T_c$ superconductivity induced within pristine 1T-TaS$_2$ by the proximity effect. The CDW order in the 1T-TaS$_2$ appears to play an important role firstly by coexisting with an unexpected and surprisingly high $T_c$ of the proximity gap in the 1T-TaS$_2$ and secondly by revealing a heretofore unseen ancillary dip-hump feature that accompanies a primary dip-hump feature corresponding to a boson (phonon) mode $\Omega$ previously seen in STM work on intrinsic Bi-2212. This second dip-hump feature is clearly related to the proximity of the CDW dominated 1T-TaS$_2$ and implies one of two possible boson modes depending on whether the dip-hump feature is referenced to the reduced gap $\Delta_p$ or the intrinsic gap $\Delta_0$ at the interface. In the former case the temperature-dependent boson mode $\Omega_p = \rm{E^{*}_p} - \Delta_p$ has values in the range 30-45 meV, somewhat less than the 40-50 meV range shown in Fig.~\ref{figure5}(c) for $\Omega$. The energy $\Omega_p$ can probably be interpreted as the mode relating to the gap $\Delta_p$ in the same manner as the mode $\Omega$ is related to the intrinsic gap $\Delta_0$. In the latter case however the temperature dependent boson mode $\Omega^* = \rm{E^{*}_p} - \Delta_0 $ has values in the range 20-25 meV which is close to the infrared active phonon mode associated with the CDW in 1T-TaS$_2$ as measured by infrared reflectance\cite{Tanner TaS2}. This more plausible latter interpretation provides independent evidence that the phonon associated with the CCDW phase coexists with and may even enhance the superconductivity in pristine 1T-TaS$_2$. However, both of the above scenarios describing the second dip--hump structure provide reasonable rationales for the occurrence of a high-$T_c$ proximity effect in Bi-2212/1T-TaS$_2$ junctions. Our work posits a mutual interaction of the CCDW and superconducting order parameters in the interfacial region of Bi-2212/1T-TaS$_2$ contacts, thereby revealing rich phenomenology and confirming a strong interplay between high $T_c$ superconductivity and CDW order which is only beginning to be understood.

\section*{Methods}

\subsection*{Sample fabrication}
\begin{figure}
\includegraphics[width=0.7\textwidth]{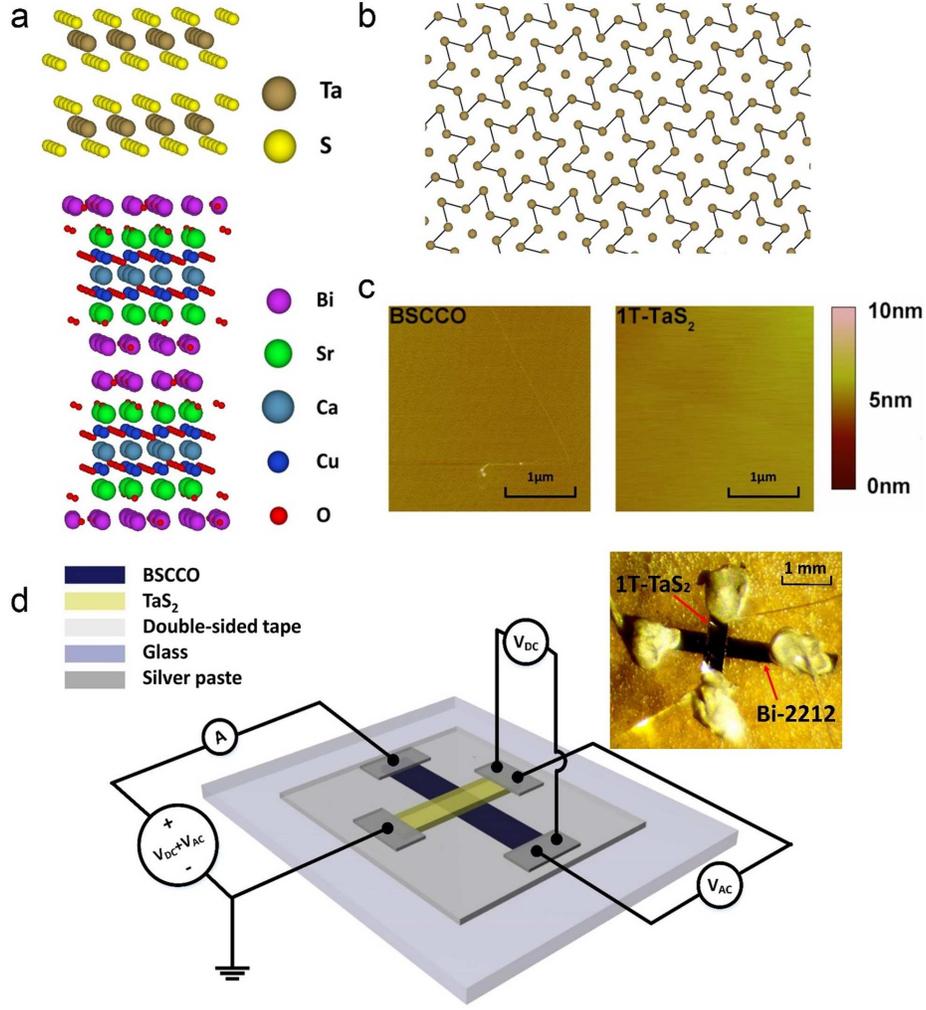}
\caption{Sample characteristics and measurement set-up. (a): Crystal structure of 1T-TaS$_{2}$ (upper schematic) and Bi-2212 (lower schematic) viewed along the direction parallel to the \textit{ab}-plane. (b):  Schematic of monolayer 1T-TaS$_{2}$ along the \textit{c}-axis in CCDW state. The interlocked clusters of Ta atoms (``star of David") are sketched in dark yellow for the Ta atoms and the chemical bonds between Ta atoms, excluding the central one, are sketched in black. The S atoms are not shown. (c): Atomic force microscope (AFM) images of the cleaved surfaces for thin flakes of Bi-2212 (left) and 1T-TaS$_{2}$ (right). The scale bar corresponds to 1 $\rm{\mu m}$. The surface mean roughness of thin flake Bi-2212 and 1T-TaS${_2}$ are 1.55\thinspace $\rm{\AA}$ and 1.52\thinspace $\rm{\AA}$ respectively. (d): Schematic depicting experimental set-up for making four-terminal current-voltage and differential conductance vs voltage measurements using an AC+DC adder in conjunction with one DC voltmeter and two synchronized lock-in amplifiers, one for AC current and the other for AC voltage. Inset: photograph of Bi-2212/1T-TaS$_2$ device. The scale bar corresponds to 1\thinspace mm.}
\label{figure1}
\end{figure}

\begin{figure}
\includegraphics[width=0.7\textwidth]{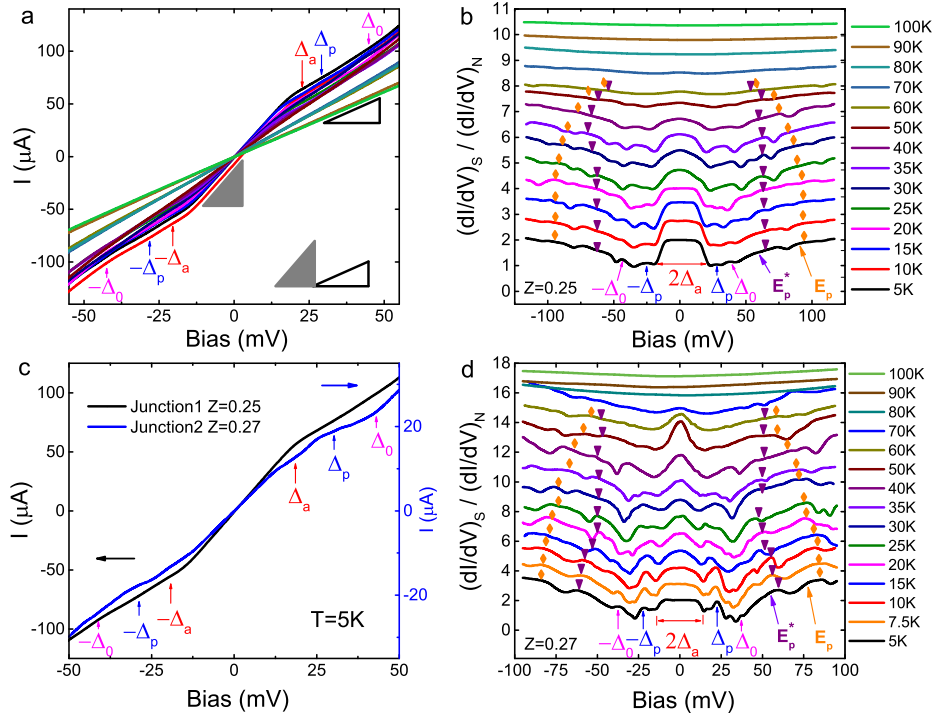}
\caption{Highly transparent Bi-2212/1T-TaS$_{2}$ junctions. (a): DC $I-V$ curves of Junction-1, with BTK parameter Z=0.25, at various temperatures from 5\thinspace K to 100\thinspace K. $\Delta_{a}$ (Red), $\Delta_{p}$ (Blue) and $\Delta_{0}$ (Magenta) respectively represent the induced superconducting gap on 1T-TaS$_{2}$ side at interface, the depressed superconducting gap on Bi-2212 side at interface, and the intrinsic superconducting gap of Bi-2212. The hollow solid triangle indicates the normal current of the junction, where the normal conductance at low temperatures in the superconducting state is approximately equal to that of the normal state. The solid gray triangle indicates the excess current due to Andreev reflection at N-S interface, where the conductance within the Andreev reflection region amounts to nearly twice the normal state conductance well below $T_{c}$. (b): Normalized AC differential conductance $(dI/dV)_{S}$/$(dI/dV)_{N}$ for Junction-1, where the differential conductance at 100\thinspace K is considered as the normal state conductance. The curves are shifted for clarity. (c): DC $I-V$ curves of Junction-1 (Z=0.25, black) and Junction-2 (Z=0.27, blue) at 5\thinspace K. The induced superconducting gap $\Delta_{a}$ and excess current within $\pm\Delta_{a}$ of Junction-2 are slightly lower relative to Junction-1. The normal conductance of Junction-2, around 0.62\thinspace $\rm{\mu A/mV}$, is lower than the normal conductance around 1.85\thinspace $\rm{\mu A/mV}$ for Junction-1. (d): AC differential conductance $(dI/dV)_{S}$/$(dI/dV)_{N}$ for Junction-2, normalized by the differential conductance at 100\thinspace K. Curves are shifted for clarity. Note: for panels (b) and (d), the orange/purple markers and arrows are discussed in the text.}
\label{figure2}
\end{figure}

\begin{figure}
\includegraphics[width=0.65\textwidth]{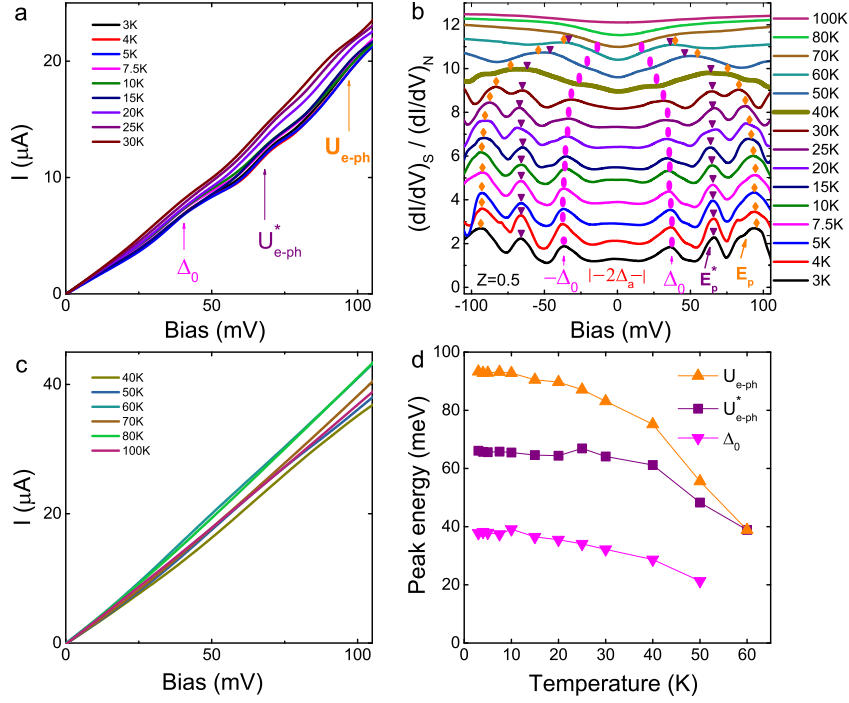}
\caption{Bi-2212/1T-TaS$_{2}$ junction with BTK Z=0.5 (Junction-3). (a): DC $I-V$ curves of Junction-3, with BTK parameter Z=0.5, at various temperatures from 3\thinspace K to 30\thinspace K, well below the $T_{c}$ of Bi-2212. The Bi-2212 intrinsic superconducting gap $\Delta_{0}$ is indicated by a magenta vertical arrow. $\rm{U}_{e-ph}$ and $\rm{U}^{*}_{e-ph}$ correspond to the peak energy positions of two nonlinear features due to the two boson modes measured with respect to $\Delta_{0}$. The normal conductance decreases with temperature, where the normal conductance at 5\thinspace K for Junction-3 is around 0.21\thinspace  $\rm{\mu A/mV}$. (b): Normalized AC differential conductance $(dI/dV)_{S}$/$(dI/dV)_{N}$ for Junction-3. The differential conductance at 100K manually scaled into the range of the conductance at one specific temperature serves as the normal conductance. Relative to the NS Andreev reflection feature seen in Junction 1 and 2, the zero-bias excess differential conductance peak is significantly reduced and almost disappears at around 40\thinspace K. The Bi-2212 intrinsic superconducting gap $\Delta_{0}$ and the humps referring to boson modes with energies $\Omega^{*}$ and $\Omega$ are indicated respectively by solid magenta ellipses, inverted purple triangles and orange diamonds. Positions of $\rm{E_{p}}$ and $\rm{E^{*}_{p}}$ are marked by orange and purple arrows. The curves are shifted for clarity. (c): DC $I-V$ curves of Junction-3 at various temperatures from 40\thinspace K to 100\thinspace K. The conductance in the normal state at 100\thinspace K is around 0.4\thinspace $\rm{\mu A/mV}$. The temperature dependence of the conductance for Junction-3 evolves consistently with intrinsic 1T-TaS$_{2}$ at temperatures from 3\thinspace K to 100\thinspace K shown in Fig. S.1(b). (d): Temperature dependence of the two humps' peak positions and the Bi-2212 intrinsic superconducting gap.}
\label{figure3}
\end{figure}

\begin{figure}
\includegraphics[width=0.8\textwidth]{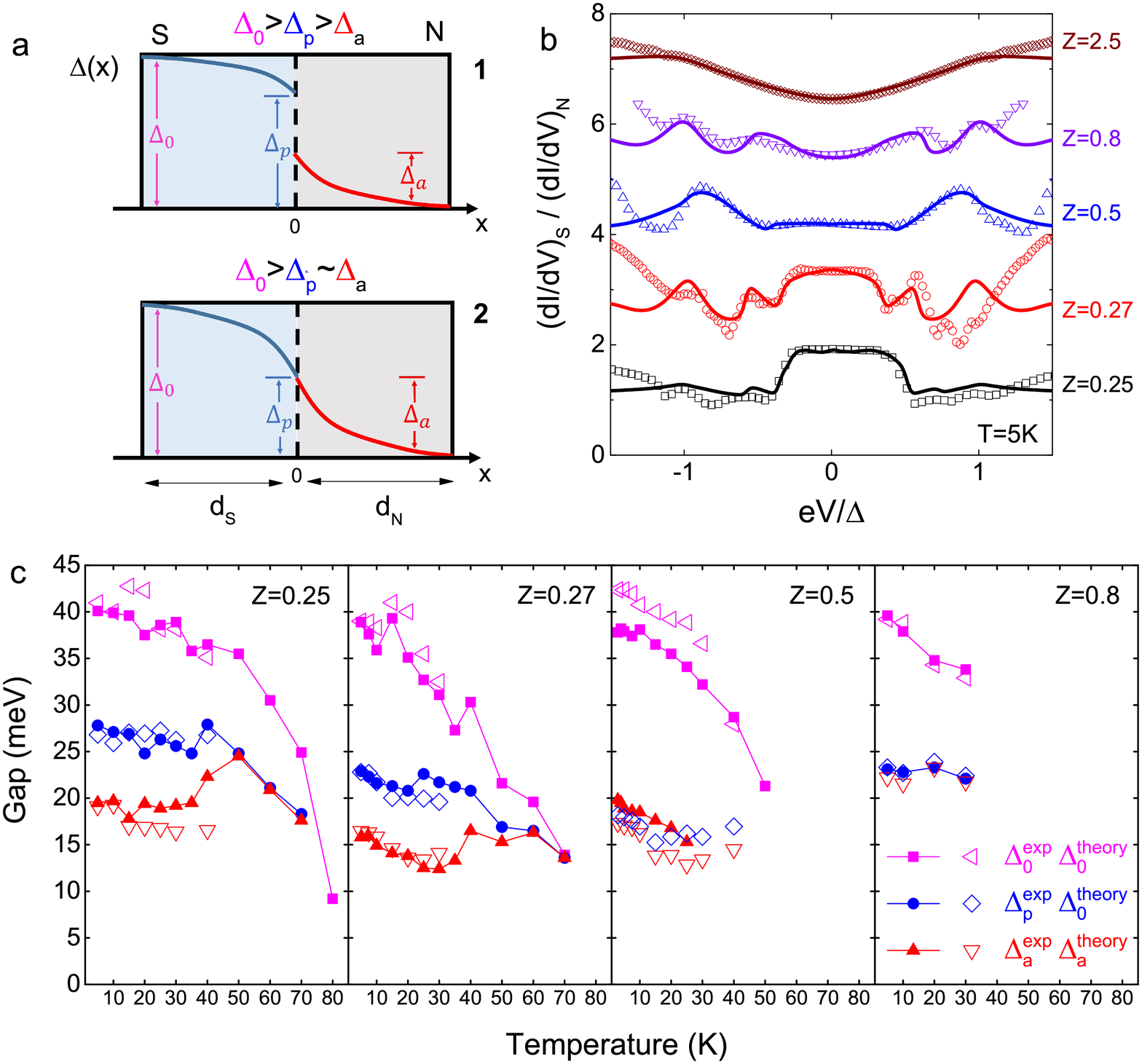}
\caption{Superconducting proximity effect and theoretical modeling. (a): Schemes of NS junction with superconducting proximity effect. $\Delta_{0}$, $\Delta_{p}$ and $\Delta_{a}$ correspond to the intrinsic superconducting gap, depressed superconducting gap on S side at the interface and induced superconducting gap on N side at the interface due to the superconducting proximity effect. Upper scheme:  $\Delta_{0}>\Delta_{p}>\Delta_{a}$; Lower scheme: $\Delta_{0}>\Delta_{p}\sim\Delta_{a}$. (b): Calculated (colored solid lines) and measured (hollow symbols) normalized differential conductance $(dI/dV)_{S}$/$(dI/dV)_{N}$ at 5\thinspace K for various Bi-2212/1T-TaS$_{2}$ junctions, with BTK parameters Z=0.25, 0.27, 0.5, 0.8 and 2.5. The junction with Z=2.5 does not show any superconducting proximity effect at the NS interface. Curves are shifted for clarity. (c): Temperature dependence of $\Delta_{0}$, $\Delta_{p}$ and $\Delta_{a}$ for various Bi-2212/1T-TaS$_{2}$ junctions: Junction-1, Z=0.25; Junction-2, Z=0.27; Junction-3, Z=0.5; Junction-4, Z=0.8. More details are discussed in the text. The gaps used in the theoretical calculation of $(dI/dV)_{S}$/$(dI/dV)_{N}$ are marked as large-sized hollow symbols, and the gaps from measurements are marked as small-sized solid symbols.}
\label{figure4}
\end{figure}

\begin{figure}
\includegraphics[width=0.78\textwidth]{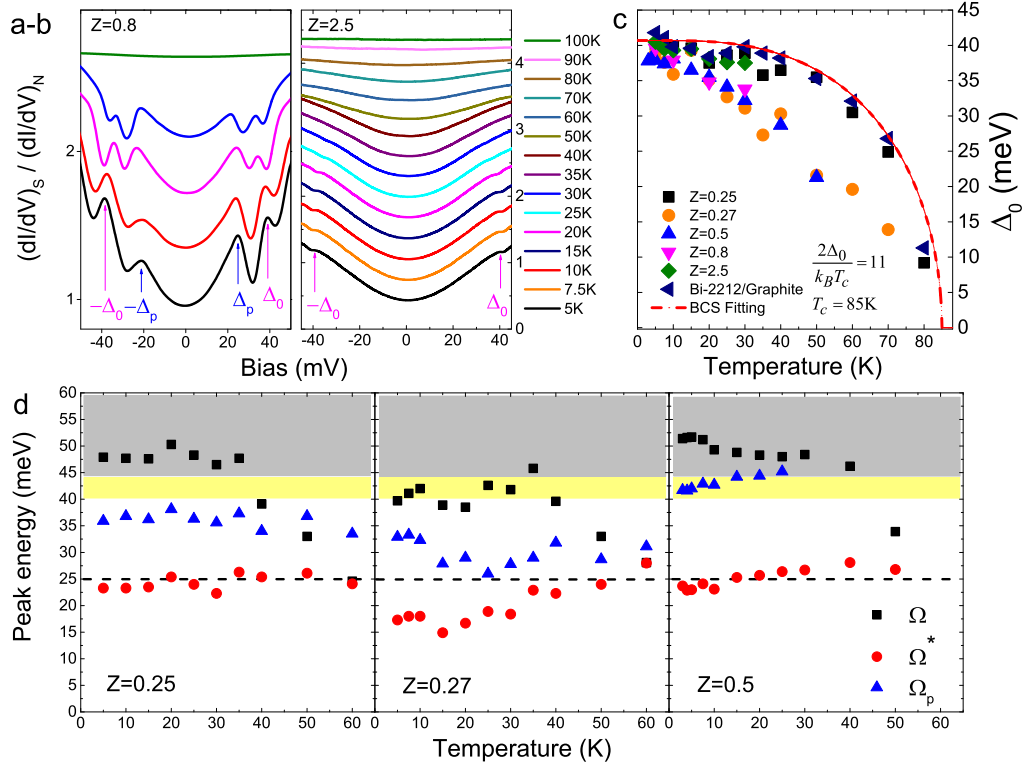}
\caption{Junctions with lower transparencies, measured Bi-2212 intrinsic superconducting gap and boson mode energies in junctions. (a-b): Normalized differential conductance spectroscopies $(dI/dV)_{S}$/$(dI/dV)_{N}$ at various temperatures (5\thinspace K to 100\thinspace K) for Junction-4 (left, BTK Z=0.8) and Junction-5 (right, BTK Z=2.5). Curves are shifted for clarity. (c): Temperature dependent Bi-2212 intrinsic superconducting gap $\Delta_{0}$ measured from various junctions, including Bi-2212/1T-TaS$_{2}$ and Bi-2212/graphite junctions. The red dashed line indicates the BCS fitting with a gap ratio $2\Delta_{sc}/k_{B}T_{c}=11$. (d): Temperature dependence of boson mode energy $\Omega=\rm{E_{p}}-\Delta_{0}$, $\Omega^{*}=\rm{E^{*}_{p}}-\Delta_{0}$ and $\Omega_{p}=\rm{E^{*}_{p}}-\Delta_{p}$ for Junction-1--3. At 5\thinspace K, $\Omega$, $\Omega^{*}$ and $\Omega_{p}$ are: 48\thinspace meV, 24\thinspace meV and 36\thinspace meV for Junction-1; 40\thinspace meV, 17\thinspace meV and 33\thinspace meV for Junction-2; 51\thinspace meV, 24\thinspace meV and 41\thinspace meV for Junction-3. Gray background indicates the energy range from 44\thinspace meV to 60\thinspace meV, yellow background indicates the energy range from 40\thinspace meV to 44\thinspace meV, and black dashed line indicates the energy level of 25\thinspace meV. Note, the $\Omega_{p}$ for Junction-3 is calculated by the predicted value based on the modeling due to the difficulty of distinguishing $\Delta_{p}$ and $\Delta_{a}$ in Junction-3.}
\label{figure5}
\end{figure}

\begin{figure}
\includegraphics[width=0.9\textwidth]{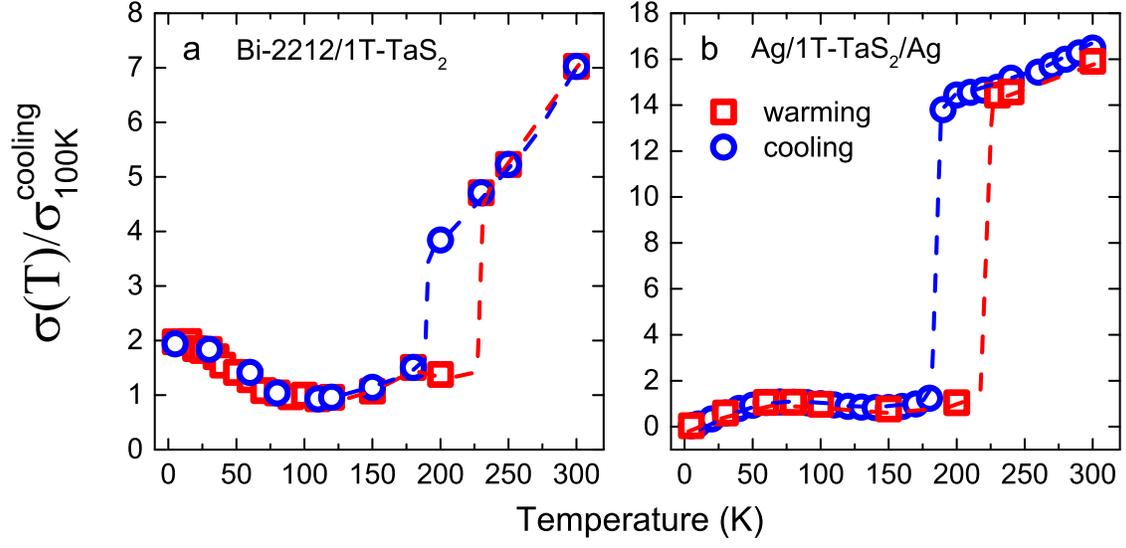}
\caption{Normalized zero-bias conductance for (a) Bi-2212/1T-TaS$_2$ Junction-1 (Z=0.25) and (b) Ag/1T-TaS$_2$/Ag junction plotted as a function of temperature under a cooling/warming cycle. The reference conductance is the zero-bias conductance at 100\thinspace K on the cooling cycle. The red and blue dashed lines indicate the CCDW-NCCDW transition of 1T-TaS$_2$.} 
\label{figure6}
\end{figure}

High quality, optimally doped single crystals of Bi$_{2}$Sr$_{2}$CaCu$_{2}$O$_{8+\delta}$ (Bi-2212) were synthesized using the method of Mitzi \textit{et al}\cite{Mitzi}, with the modification that a Pt crucible was used in place of an alumina one. This avoided possible contamination of the melt via reaction with the crucible walls. Single crystals of 1T-TaS${_2}$ flakes were prepared using iodine vapor transport\cite{vapor trans}. The transport measurements shown in Fig. S 1 verify the critical superconducting temperature of Bi-2212 to be 85\thinspace K, and the CCDW-NCCDW transition temperatures for cooling and warming processes of 1T-TaS$_{2}$ to be 180\thinspace K and 230\thinspace K respectively. 
Prior to junction fabrication, thick flake Bi-2212 (thickness around 10\thinspace $\mu m$ was mechanically exfoliated via Nitto-REVALPHA thermal release tape. Then, thick flake Bi-2212 was transferred to a clean glass substrate by thin double sided tape. Thin flake Bi-2212 (0.5 to 2\thinspace $\mu m$) was then cleaved via Scotch tape. A cleaved thin flake of 1T-TaS$_{2}$ (2 to 5\thinspace $\mu m$) was immediately placed against the cleaved surface of Bi-2212 after similar mechanical exfoliation. The two cleaved flakes strongly adhere to each other via Van der Waals forces, naturally forming high quality NIS or NS heterostructures. Junctions with different transparencies, characterized with theoretical modeling by the extended BTK model described in the manuscript, were randomly achieved. No significant dependence of the transparency upon area of junction or thickness of flake was recognized. All steps of fabrication were performed in a dry atmosphere, and all cleaved surfaces of thin flake Bi-2212 and 1T-TaS$_{2}$ were clean and flat. Based on multiple measured local-areas on the exfoliated pieces, the mean roughness for both materials is within 2\thinspace $\rm{\AA}$ as indicated by the AFM images shown in Fig.~\ref{figure1}(c)). Also, to implement four-terminal \textit{c}-axis differential conductance measurements, the thin flake Bi-2212 and 1T-TaS$_{2}$ were fashioned into narrow rectangular shapes oriented perpendicular to each other with an overlap area around 0.2\thinspace $\rm{mm^{2}}$, as shown in Fig.~\ref{figure1}(d).

\subsection*{Experimental measurements set-up}
All measurements including the four-terminal AC differential conductance measurements and the DC current-voltage ($I-V$) measurements were performed over a wide range of temperature (2.5\thinspace K to 120\thinspace K) using a Quantum Design Physical Properties Measurement System (PPMS). Samples were mounted on a commercial PPMS puck and all measurements were performed in a low-noise screen room. For AC measurements, we used 23.3\thinspace Hz as the AC output frequency. The DC source voltage was supplied by a Keithley 2400 source meter and the AC source was supplied by a Agilent 33120A AC generator with $\Delta V\sim0.2\thinspace \rm{mV}$. The DC and AC source signals were added using a homemade DC+AC adder and then applied to the junction. The DC bias across the junctions was measured with HP 3456A multimeter, the AC voltage signal with a SR830 DSP lock-in amplifier, and the AC current signal with a second SR830 lock-in amplifier after converting the current to a voltage using a SR570 current preamplifier.

\subsection*{Data normalization}
For non-ideal NS junctions which have different density of states on the N and S sides, the conductance at one specific temperature cannot be simply normalized because of temperature dependence of the normal conductance. Thus, we normalized the AC differential conductance of non-ideal NS junctions, such as Junction-3, Junction-4 and Junction-5 discussed in the manuscript, by using the conductance at 100\thinspace K manually scaled into the range of the normal conductance at one specific temperature as the effective normal conductance. For instance,
in Junction-3 the normal conductance at 5\thinspace K is around 0.2\thinspace $\rm{\mu A/mV}$ and the conductance of the normal state at 100K is around 0.4\thinspace $\rm{\mu A/mV}$. Hence, the ratio factor $r_{eff}$ is the ratio of the normal conductance at 5\thinspace K to conductance at 100\thinspace K, which is around 0.5. Then, the normalized differential conductance spectroscopy at 5\thinspace K can be determined by multiplying a ratio factor $r_{eff}$ when calculating the division of $(dI/dV)_{S}$ by $(dI/dV)_{N}$.

\section*{Acknowledgements}
The authors thank C. Samouce for the AFM measurements, A. G. Rinzler for supporting the related AFM facility and P. J. Hirschfeld for useful discussions. This work was supported by the National Science Foundation under Grant No. DMR--1305783 (AFH) and the Department of Energy under Grant No. DE-FG02-86ER45268 (GRS).

\section*{Author contributions}
A. J. L and A. F. H conceived and planned the research. A. J. L and X. Z implemented the measurement set-up and device fabrication. G. R. S synthesized the high quality crystals of Bi-2212 and 1T-TaS$_{2}$. A. J. L performed the measurements. X. Z helped with the measurements. A. J. L carried out the theoretical modeling and calculation. A. J. L and A. F. H wrote the paper with input from all authors. A. F. H supervised the work. All the authors contributed and agreed to the scientific discussion and manuscript revision.

\newpage

\newpage

\newpage

\newpage

\newpage

\newpage

\clearpage
\section*{Supplementary Information}

\setcounter{figure}{0}
\renewcommand{\figurename}{Supplementary Fig S.}

\section*{Characterization of Intrinsic Crystals }
\subsection*{Experimental Method}
Prior to transport measurement, single crystal Bi-2212 and 1T-TaS$_{2}$ samples (crystal structure shown in Fig.1(a) in manuscript) were first mechanically exfoliated in a dry atmosphere to produce thick flakes with thickness around 10\thinspace $\rm{\mu m}$. Then, four-terminal transport measurements (contact configuration is shown in Fig S.~\ref{Supp figure1}(a) inset) were carried out using a SR830 DSP Lock-in Amplifier with an AC voltage of 0.5\thinspace V, frequency of 526\thinspace Hz and input impedance of 100\thinspace $\rm{k\Omega}$. The four-terminal configuration was constructed by gold wires and SPI silver paste. The temperature range for transport measurements is up to 300\thinspace K.

\subsection*{Bi-2212}
Nearly optimal-doped high-$T_{c}$ crystals of cuprate Bi-2212 (Bi$_{2}$Sr$_{2}$CaCu$_{2}$O$_{8+x}$) were used in this work. The critical superconducting temperature $T_{c}$ in bulk Bi-2212 was verified to be 85\thinspace K via transport measurement as shown in Fig S.~\ref{Supp figure1}(a). The Bi-2212 intrinsic superconducting gap via our measurements (see details in manuscript on Bi-2212/1T- TaS$_{2}$ junctions and supplementary information on Bi-2212/graphite junction below) is around 38-42\thinspace $\rm{meV}$ at 5\thinspace K, which reveals the BCS gap ratio $2\Delta_{sc}/k_{B}T_{c}$ to be around 10.4-11.5, in good agreement with previous works on the intrinsic Bi-2212 superconducting gap\cite{Fischer BSCCO gap, Fischer BSCCO gap1}.

\subsection*{1T-TaS$_{2}$}
\begin{figure}
\includegraphics[width=0.8\textwidth]{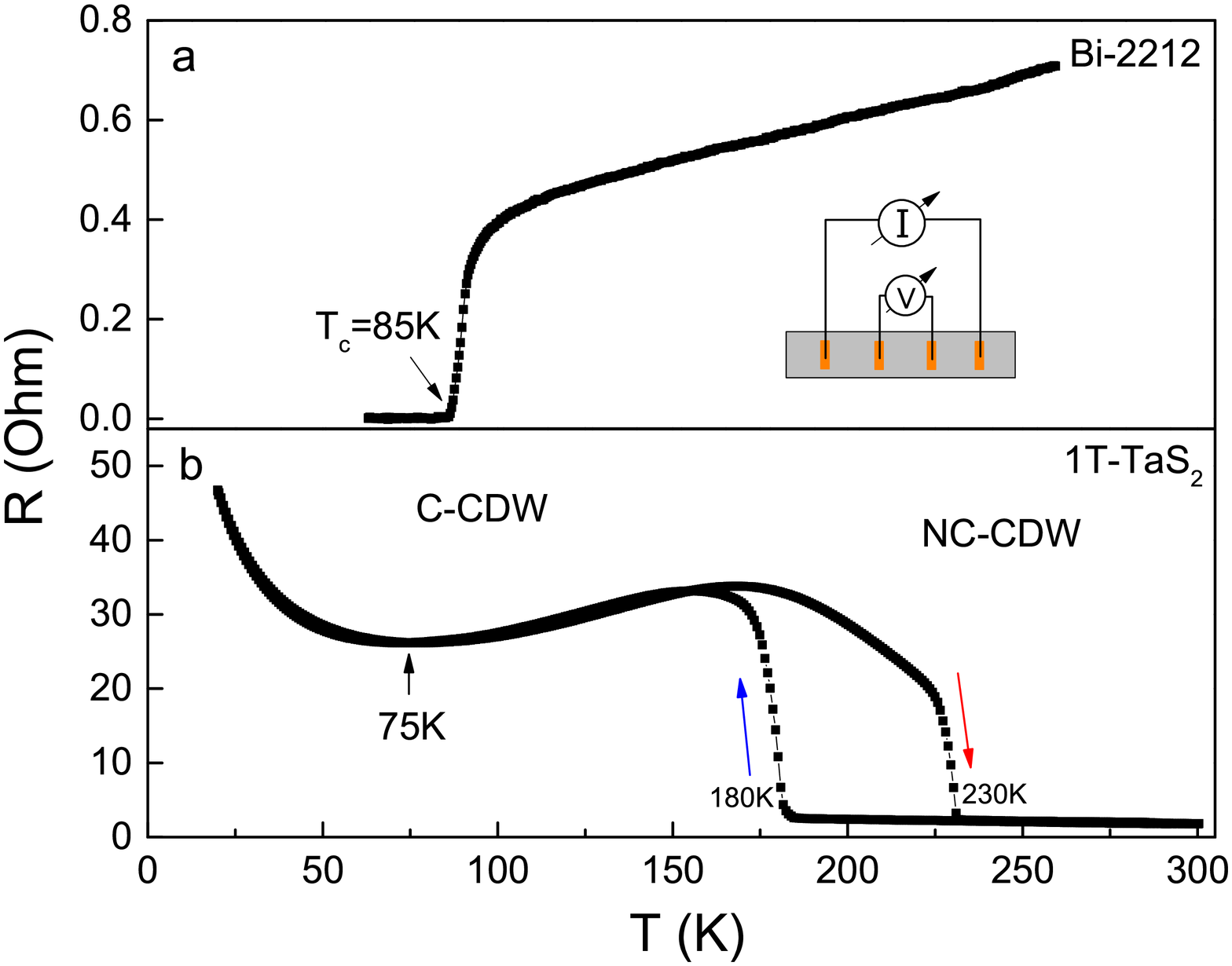}
\caption{Intrinsic crystal characteristics. (a): Zero field AC four-terminal transport measurement of Bi-2212 single crystalline flake, indicating the superconducting critical temperature at 85\thinspace K. (Inset: Schematic configuration of four-terminal transport measurement.) (b): Zero field AC four-terminal transport measurement of 1T-TaS$_{2}$ single crystalline flake. Upward blue and downward red arrows indicate the critical temperatures of CCDW-NCCDW phase transition at cooling and warming process respectively. The vertical arrow indicates the temperature near 75\thinspace K where the resistivity begins to increase with decreasing temperature.}
\label{Supp figure1}
\end{figure}

The 1T-TaS$_{2}$ used in our Bi-2212/1T-TaS$_{2}$ junctions belongs to the layered transition metal dichalcoginide (TMD) family and has a series of CDW states persisting over a wide temperature range. The thin-flake 1T-TaS$_{2}$ transport measurement (shown in Fig. S.~\ref{Supp figure1}(b)) reveals with decreasing temperature a transition from a nearly commensurate CDW state (NCCDW) to a commensurate CDW state (CCDW)\cite{TaS2 phase transition}. The CDW state is a highly ordered configuration in which the central Ta atom is surrounded by 12 Ta atoms forming a ``star of David'' array\cite{TaS2davidstar}) which appears near 180\thinspace K on cooling and vanishes near 230\thinspace K on warming. As temperature decreases, the resistance increases and transport is dominated by a Mott-CCDW ground state in which the Mott insulating state coexists (or resides in) the CCDW state. The upturn in resistivity starting at around 75\thinspace K indicates the localization of electrons due to a strong electron-electron Coulomb interaction considered to be the mechanism for the formation of the Mott-insulating state\cite{MI1, MI2, MI3, MI4}.
To verify the intrinsic superconducting gap of thin flake Bi-2212 (thickness around 0.5 to 5\thinspace $\rm{\mu m}$) along \textit{c}-axis, we measured several Bi-2212/graphite junctions by replacing the 1T-TaS$_2$ with graphite and using the same technique as discussed in the manuscript.  The freshly exfoliated highly ordered pyrolytic graphite (HOPG) flake is a van der Waals electrode with high conductivity. Consequently, a good normal metal-insulator-superconductor (NIS) or normal metal-superconductor (NS) junction is naturally formed. 

\section*{Bi-2212/Graphite Junction}

\begin{figure}
\includegraphics[width=0.75\textwidth]{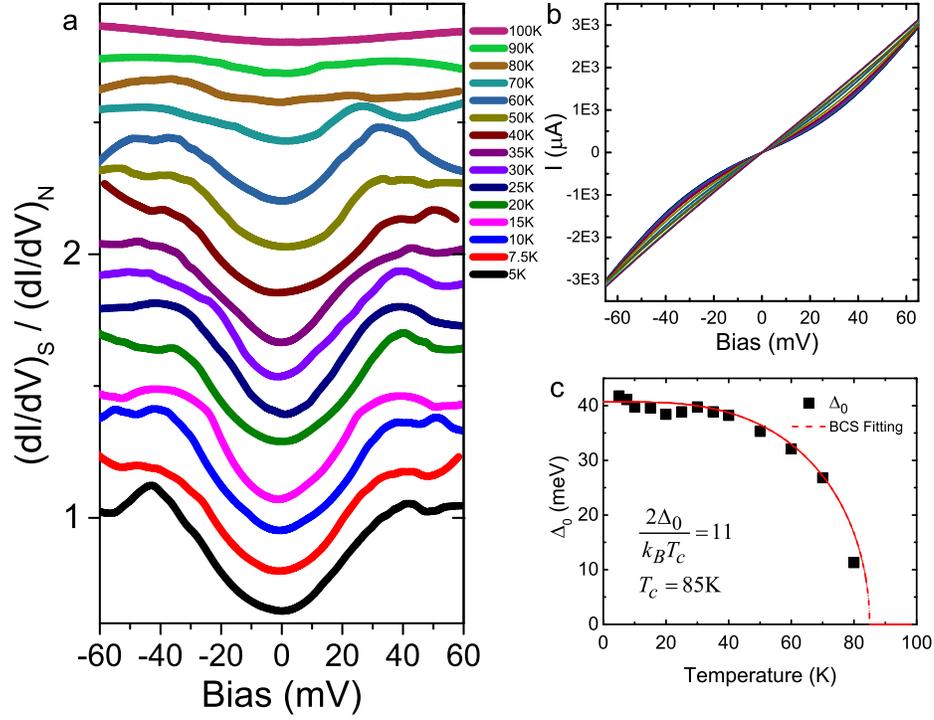}
\caption{Bi-2212/graphite junction measurements. (a): The AC differential conductance normalized by the differential conductance at 100\thinspace K, $(dI/dV)_{S}/(dI/dV)_{N}$, for various temperatures from 5\thinspace K to 100\thinspace K. The colored isotherms with temperatures identified in the legend of panel (a) are shifted for clarity. (b): DC $I-V$ curves at selected temperatures related to the AC differential conductance measurement shown in panel (a). (c): Temperature dependence of the intrinsic superconducting gap $\Delta_{0}$ of thin flake Bi-2212, experimentally determined  from the tunneling junction with graphite (black squares). The red dashed line indicates the best fit using the BCS gap function with the strong coupling ratio $2\Delta_{0}/k_{B}T_{c}=$11. }
\label{Supp figure2}
\end{figure}

As shown in Fig.~S.~\ref{Supp figure2}(a), the differential conductance $(dI/dV)_{S}$, normalized by the normal state conductance $(dI/dV)_{N}$ at 100\thinspace K, shown in Fig.~S.~\ref{Supp figure2}(b), clearly shows the commonly accepted \textit{c}-axis density of states (DOS) of $d$-wave superconductors without any evidence for a superconducting proximity effect. At low temperatures, the superconducting gap measured from the Bi-2212/graphite junction is around 40\thinspace meV. This result not only shows good agreement of the intrinsic gap of Bi-2212 with previous tunneling spectroscopy studies\cite{Fischer BSCCO gap, Fischer BSCCO gap1}, but also reveals consistency with the Bi-2212/1T-TaS$_{2}$ junctions discussed in the manuscript. In addition, with a BCS gap function ratio $2\Delta_{0}/k_{B}T_{c}=$11 and $T_{c}=85\thinspace \rm{K}$, we find for our samples the measured temperature dependence of the superconducting gap is well described by the BCS gap function\cite{BCS gap} as shown in Fig.~S.~\ref{Supp figure2}(c).

\section*{Theoretical Modeling}
This section is mainly focused on the theoretical calculation for understanding the density of states (DOS) features at Bi-2212/1T-TaS$_{2}$ interface with the effect of superconducting proximity at an energy scale within the intrinsic superconducting gap. The discussion of the multiple dip-hump structures in the manuscript will not be treated in a more quantitative way, since there is no theoretical consensus on the physical mechanisms of peak-dip-hump structures seen in high-T$_{c}$ cuprates\cite{diphump model1, diphump model2, diphump model3, diphump model4}. 

For normal metal-insulator-superconductor (NIS) junctions, the conductance spectrum is calculated using two reflection rates: ordinary reflection (OR) in which incident electrons are reflected as electrons  $R_{ee}$ and Andreev reflection (AR) in which incident electrons are reflected $R_{eh}$ as holes. These parameters, $R_{ee}$ and $R_{eh}$, appear in the Bogoliubov-de Gennes  (BdG) equations and are used in the formula of Blonder-Tinkham-Klapwidjk (BTK) on the normal material side of NIS (or NS) junction\cite{BTK} to describe the current

\renewcommand{\theequation}{S.\arabic{equation}}
\be
\begin{split}
I_{NS} &=2N(0)ev_{F}A\int_{-\infty}^{+\infty}[f_{\rightarrow}(E)-f_{\leftarrow}(E)]dE\\
&=2N(0)ev_{F}A\int_{-\infty}^{+\infty}[f_{0}(E-eV)-f_{0}(E)][1+R_{eh}^{2}(E)-R_{ee}^{2}(E)]dE~,
\end{split}
\label{BTK eqn}
\ee
where, A is the area of interface, $f_{0}(E)$ is the Fermi-Dirac distribution at temperature $T$, $v_{F}$ is the fermi velocity, $N(0)$ is the one-spin density of states at $E=E_{F}$ in the normal state when $N_{S}(0)=N_{N}(0)$, $E<<E_{F}$ for junctions, and
\be
\begin{split}
f_{\rightarrow}(E)&=f_{0}(E-eV)\\
f_{\leftarrow}(E)&=R_{eh}(E)f_{0}(E+eV)+R_{ee}f_{0}(E-eV)+[1-R_{eh}(E)-R_{ee}(E)]f_{0}(E)~.
\end{split}
\label{BTK f}
\ee
Accordingly, the differential conductance $dI/dV$ is expressed as
\be
\frac{dI}{dV}(V)=2N(0)ev_{F}A\int_{-\infty}^{+\infty}\frac{\partial f_{0}(E-eV)}{\partial (eV)}[1+R_{eh}(E)-R_{ee}(E)]dE~.
\label{didv0}
\ee

In more complicated cases instead of an ideal thin junction, there is a superconducting proximity effect in junctions with larger thickness than the superconducting coherence length because of the thick boundary or interface\cite{deGennes proximity, McMillan}. The thicknesses of the proximity region of S and N are $d_{S}$ and $d_{N}$ respectively. The schemes of superconducting proximity effect are shown in Fig.~4(a) in the manuscript.

However, the original BTK model\cite{BTK} depicts the tunneling spectrum in the case of conventional superconductors where the superconducting gap is isotropic in momentum space. For unconventional superconductors, with anisotropic gap symmetry in momentum-space, such as $d$-wave, $p$-wave, $s+d$ wave, etc, the BTK model needs to be modified. Since Bi-2212 is a well-known high-T$_{c}$ cuprate superconductor with $d$-wave gap symmetry in momentum-space\cite{bsccodwave1,bsccodwave2,bsccodwave3,bsccodwave4,bsccodwave5}, we start our theoretical calculation based on the previous works of the extended BTK model on tunneling spectrum for $d$-wave unconventional superconductors \cite{BTKhtc1, BTKhtc2, BTKhtc3}. 

The BdG equations for unconventional superconductors with the momentum dependent pairing potential $\Delta(\gamma, r)$ are expressed as
\be
\begin{split}
Eu(\gamma,r)&=H_{0}(r)u(\gamma,r)+\Delta(\gamma,r)v(\gamma,r)\\
Ev(\gamma,r)&=-H_{0}(r)v(\gamma,r)+\Delta^{*}(\gamma,r)u(\gamma,r)~,
\end{split}
\label{BdG}
\ee
where $\gamma=\frac{k}{k_{FS}}$; $u(\gamma,r)$ and $v(\gamma,r)$ are the solutions of the BdG equations for electron-like (ELQ) and hole-like (HLQ) quasiparticles. The Hamiltonian part is $H_{0}(r)=-\hbar^{2}\nabla^{2}_{r}/2m-\mu+V(r)$. 

For \textit{c}-axis tunneling in $d$-wave superconductors\cite{BTKhtc1}, the ELQ (+) and HLQ (--) are experiencing the same pairing potential magnitude ($|\Delta_{+}|=|\Delta_{-}|=\Delta_{0}\rm{cos}(2\alpha)$, $\alpha$ is the angle of one specific orientation away from the lobe of gap in momentum-space), and global phase ($\phi_{+}=\phi_{-}=0$). So, after solving the BdG equations with boundary conditions\cite{BTKhtc1,BTKhtc2,BTKhtc3}, the AR rate $R_{eh}(E)$ and OR rate $R_{ee}(E)$ at energy $E$ away from $E_{F}$ are found to be
\be
\begin{split}
R_{eh}(E)&=\frac{e^{-i\theta_{+}}\sqrt{E+\Omega_{-}}\sqrt{E-\Omega_{+}}}{(1+Z^{2})\sqrt{E+\Omega_{-}}\sqrt{E+\Omega_{+}}-e^{i(\theta_{d}+\theta_{-}-\theta_{+})}Z^{2}\sqrt{E-\Omega_{-}}\sqrt{E-\Omega_{+}}}\\\\
R_{ee}(E)&=\frac{-Z(i+Z)[\sqrt{E+\Omega_{-}}\sqrt{E
+\Omega_{+}}-e^{i(\theta_{d}+\theta_{-}-\theta_{+})}\sqrt{E-\Omega_{-}}\sqrt{E-\Omega_{+}}]}{e^{2iq^{+}d_{N}}(1+Z^{2})\sqrt{E+\Omega_{-}}\sqrt{E+\Omega_{+}}-e^{i(\theta_{d}+\theta_{-}-\theta_{+})}Z^{2}\sqrt{E-\Omega_{-}}\sqrt{E-\Omega_{+}}}~.
\end{split}
\label{dwaveBTK}
\ee
Here, $\Omega_{\pm}=\sqrt{E^{2}-\Delta_{0}^{2}}$, $Z=\frac{2mH}{\hbar^{2}k_{F}\rm{cos}\theta}$, $q^{+}=\sqrt{k_{F}^{2}+2mE/\hbar^{2}}\rm{cos}\theta$, $\theta$ is the angle of the incident orientation of injected electrons relative to the normal orientation of interface, and $Z$ is the well-known BTK parameter representing the dimensionless normal conductance or transparency of interface $\sigma_{N}=\frac{4\lambda}{(1+\lambda)^{2}+4Z^{2}}$, where the parameter $\lambda$ describes the mismatch ratio of interface $\lambda=k_{FS}/k_{FN}$. For simplicity, we set $k_{FN}=k_{FS}$ resulting in equations.~(\ref{dwaveBTK}). 

In addition, the parameter $\theta_{d}$ in equations.~(\ref{dwaveBTK}) is a parameter corresponding to the thicknesses of the proximity regions. For the case in which only the N side has proximity effect, as proposed in references\cite{BTKhtc1,BTKhtc2,BTKhtc3}, $\theta_{d}=\frac{4md_{N}E}{\hbar^{2}k_{F}\rm{cos}\theta}$. If we use the BCS coherence length of intrinsic superconductor $\xi_{0}=\hbar v_{F}/\pi\Delta_{0}$ then $\theta_{d}=\frac{4}{\pi \rm{cos}\theta}\frac{d_{N}}{\xi_{0}}\frac{E}{\Delta_{0}}$\cite{BCScoherence}. 

However, generically, since there is a superconducting proximity effect on both sides of S and N\cite{deGennes proximity, McMillan}, we can slightly modify the parameter to be different within three regions, $-d_{S}<x<0_{-}$, $0_{-}<x<0_{+}$ and $0_{+}<x<d_{N}$:
\be
\theta_{d}(x)=
\begin{cases}
\frac{4}{\pi \rm{cos}\theta}\frac{d_{N}}{\xi_{0}}\frac{E}{\Delta_{0}} & 0_{+}<x<d_{N}\\
\frac{4}{\pi \rm{cos}\theta}\frac{d_{vdW}}{\xi_{0}}\frac{E}{\Delta_{0}} & 0_{-}<x<0_{+}~~,\\
\frac{4}{\pi \rm{cos}\theta}\frac{d_{S}}{\xi_{0}}\frac{E}{\Delta_{0}} &-d_{S}<x<0_{-}
\end{cases}
\label{thetacase}
\ee
where $d_{vdW}$ is the effective thickness of the Van der Waals stacking length.

With the extended tunneling model\cite{BTKhtc1,BTKhtc2}, the normalized differential conductance $\frac{(dI/dV)_{S}}{(dI/dV)_{N}}(V)$ is calculated by equation.~(\ref{didvext}) expressed below
\be
\frac{(dI/dV)_{S}}{(dI/dV)_{N}}(V)=\int_{-\infty}^{+\infty}\frac{\partial f_{0}(E-eV)}{\partial (eV)}\sigma_{T}(E)dE~,
\label{didvext}
\ee
where $\sigma_{T}(E)=\frac{\int_{\Omega}[1+R_{eh}^{2}(E)-R_{ee}^{2}(E)]\sigma_{N}\rm{cos}\theta d\Omega}{\int_{\Omega}\sigma_{N}\rm{cos}\theta d\Omega}$ and $\Omega$ refers to the semi-spherical solid angle integration over the Fermi surface of the $d$-wave superconductor. 

In addition, we consider how the thicknesses of proximity regions could affect the tunneling features, the quasiparticle life time ($\tau_{R}$) or the scattering rate ($1/\tau_{R}$).  Different quasiparticle scattering rates inside superconducting regions corresponding to the intrinsic and proximate superconducting gaps will have different smearing effects near the edge of the superconducting gap \cite{gamasmear}. Following the result by Dynes \textit{et al}, the finite quasiparticle lifetime induced smearing effect on tunneling spectroscopy is calculated via a simple modification by including an imaginary term $-i\Gamma$ in $\sqrt{E\pm \Omega_{\pm}}$ and $\Omega_{\pm}$ so that $E\rightarrow E-i\Gamma$ and $\Omega_{\pm}\rightarrow \sqrt{(E-i\Gamma)^{2}-\Delta_{0}^{2}}$. Such a method was also demonstrated by Plecenik \textit{et al.} by including an additional term $-i\Gamma$ in the Hamiltonian of the BdG equations \cite{bscco lifetime}. The term $\Gamma$ is the quasiparticle lifetime parameter, as $\Gamma=\hbar/\tau_{R}$\cite{bscco lifetime}, and $\frac{1}{\tau_{R}}=(\frac{k_{B}T}{\Delta_{0}})^{1/2}\frac{1}{\tau_{0}}e^{-\Delta_{0}/k_{B}T}$ \cite{gamasmear}, where $\tau_{0}$ is a parameter related to the electron-phonon coupling strength. As temperature increases, the quasiparticle scattering rate increases resulting in a broadening of the tunneling spectrum features.

\begin{figure}
\includegraphics[width=0.7\textwidth]{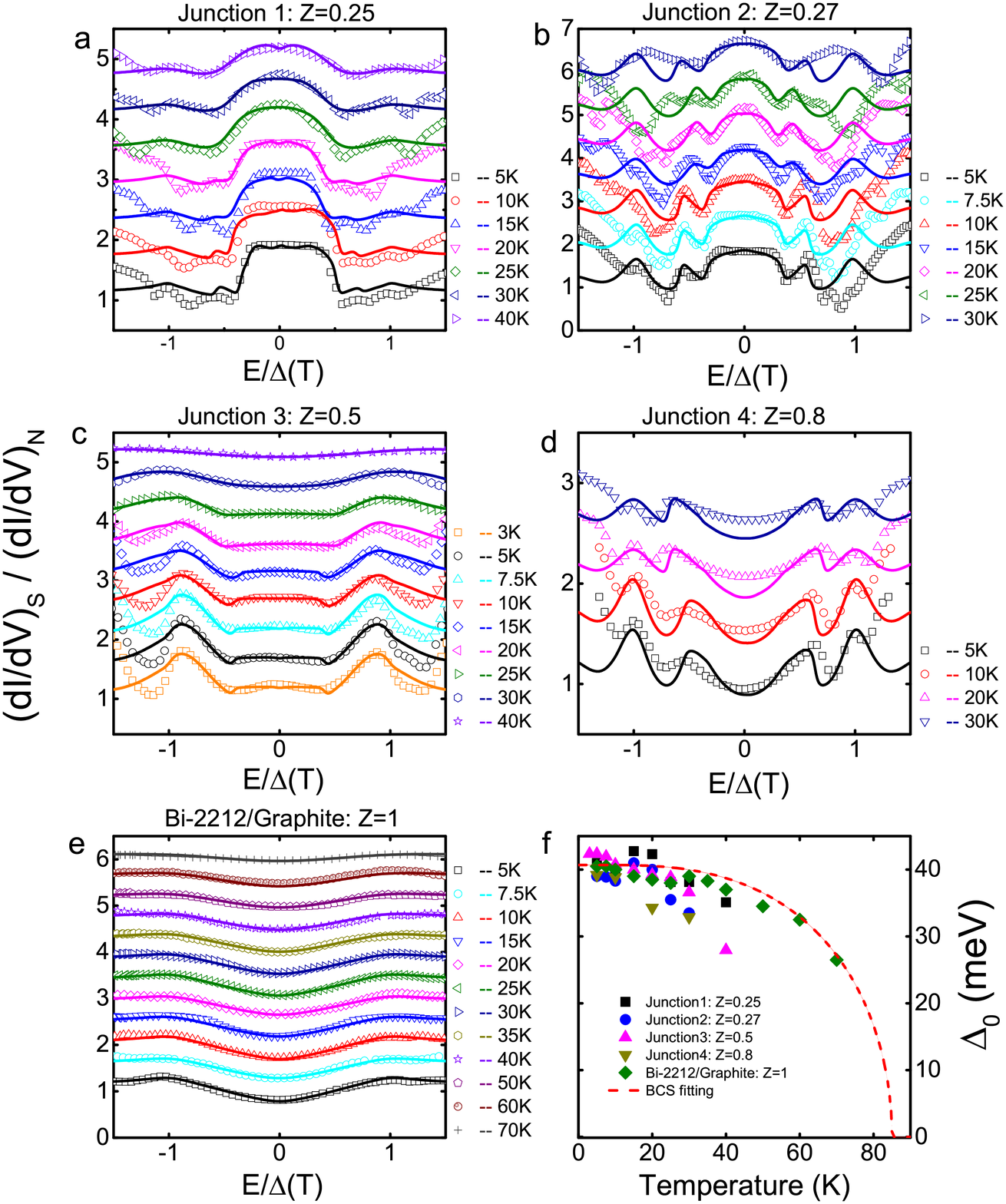}
\caption{Calculated normalized conductance $(dI/dV)_{S}/(dI/dV)_{N}$ (solid curves) compared to experimental measurement (hollow symbols). (a)--(e): Parameters used in calculations at 5K are listed in the Table\thinspace I of manuscript. Curves at different temperatures are shifted for clarity. (f): Temperature dependent $\Delta_{0}^{\rm{theory}(T)}$ used in calculation for various junctions. Red dashed line indicates the best fitting by BCS gap function with a ratio  $2\Delta_{0}/k_{B}T_{c}=$11.}
\label{Supp figure3}
\end{figure}

The calculated normalized differential conductance $(dI/dV)_{S}/(dI/dV)_{N}$ for various junctions, Bi-2212/1T-TaS$_{2}$ (Z=0.25, 0.27, 0.5 and 0.8) and Bi-2212/Graphite (Z=1), compared to the normalized conductance from measurement are shown in Fig S.~\ref{Supp figure3}. The results reveal good agreement between the theoretical model and experimental measurement. Based on the parameters used in theoretical modeling and calculation: Z is the parameter of BTK model\cite{BTK}; $\Gamma_{0}$, $\Gamma_{p}$ and $\Gamma_{a}$ are the quasiparticle lifetime parameters related to $\Delta_{0}$, $\Delta_{p}$ and $\Delta_{a}$ respectively. The parameters used in the calculation at 5\thinspace K for various junctions are listed in Table~I in the manuscript.  

\section*{Recognizing two Boson modes from $dI^{2}/d^{2}V$ spectra}

\begin{figure}
\includegraphics[width=1\textwidth]{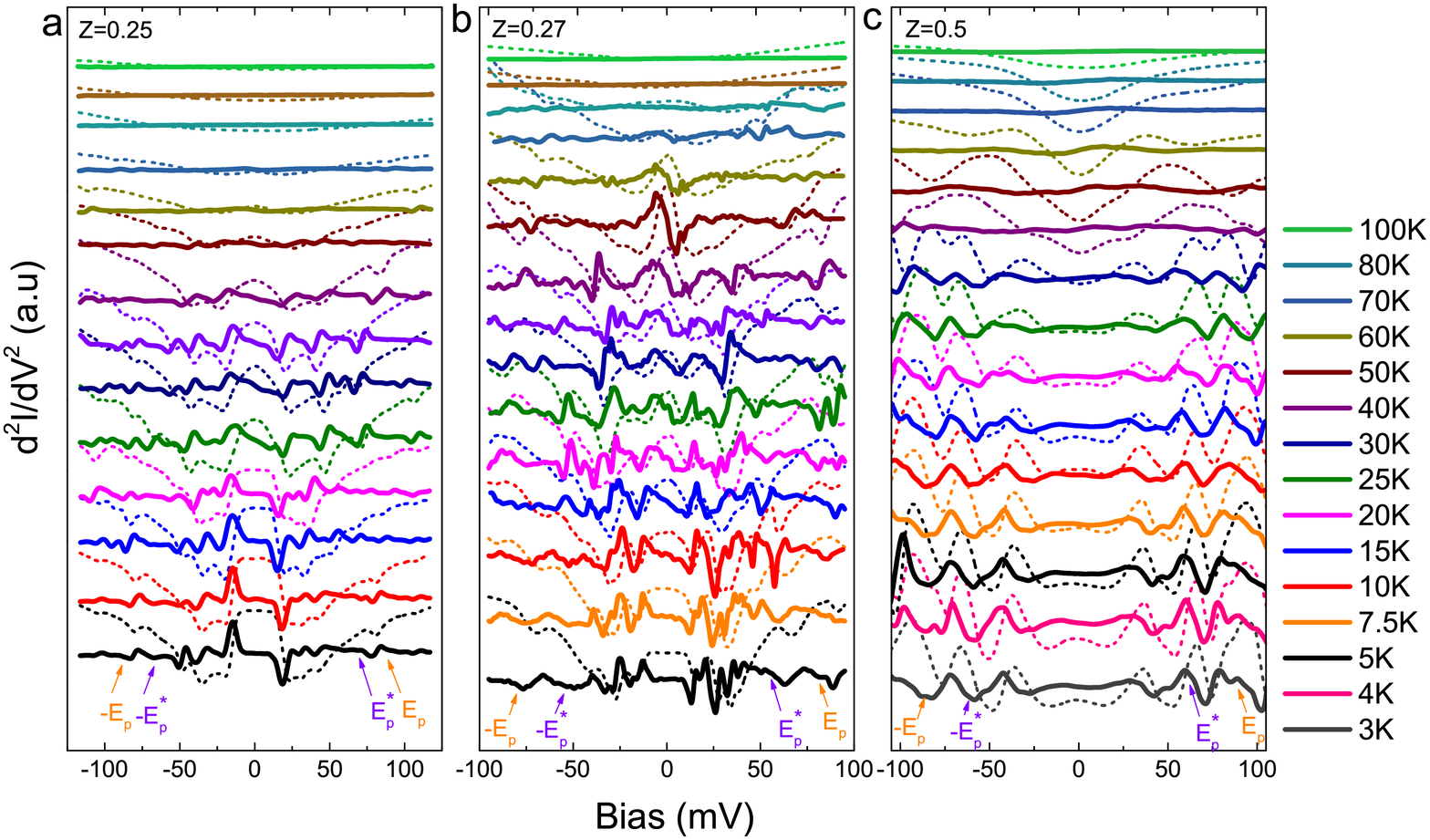}
\caption{$dI^{2}/d^{2}V$ spectrum of Bi-2212/1T-TaS$_2$ junctions.  Solid lines indicate the curves of $dI^{2}/d^{2}V$ spectrum in arbitrary units, and dashed lines indicate the normalized differential conductance $(dI/dV)_{S}/(dI/dV)_{N}$. Curves are shifted for clarity. (a): $dI^{2}/d^{2}V$ spectrum of Junction-1 (BTK Z=0.25) for various temperatures from 5\thinspace K to 100\thinspace K. (b): $dI^{2}/d^{2}V$ spectrum of Junction-2 (BTK Z=0.27) for various temperatures from 5\thinspace K to 100\thinspace K. (c): $dI^{2}/d^{2}V$ spectrum of Junction-3 (BTK Z=0.5) for various temperatures from 3\thinspace K to 100\thinspace K. $\rm{E_{p}}$ and $\rm{E^{*}_{p}}$, indicated by orange and purple arrows, correspond to the peaks position of two dip-hump structures under conductance spectroscopies in $dI^{2}/d^{2}V$ spectrum for junctions. At 5K, for Junction-1: $\rm{E_{p}}=88\thinspace \rm{meV}$ and $\rm{E^{*}_{p}}=64\thinspace \rm{meV}$; for Junction-2: $\rm{E_{p}}=79\thinspace \rm{meV}$ and $\rm{E^{*}_{p}}=56\thinspace \rm{meV}$; for Junction-3: $\rm{E_{p}}=89\thinspace \rm{meV}$ and $\rm{E^{*}_{p}}=62\thinspace \rm{meV}$.}
\label{Supp figure4}
\end{figure}

As theory in strong-coupling superconductivity describes\cite{boson mode energy0,boson mode energy1,boson mode energy2,boson mode energy3,boson mode energy4,boson mode energy5}, the impact of electron--phonon interactions on the density of states of superconductors is predicted to occur near energies $E=\Delta+\Omega$, where $\Delta$ is the superconducting gap and $\Omega$ is the corresponding boson mode energy. Such signature corresponding to the electron--phonon interaction is the dip-hump structure in the conductance spectroscopies ($dI/dV$ vs. $V$) of NIS junctions. The boson mode energy $\Omega$ is obtained by identifying the signature of a peak in the positive bias region or a dip in the negative bias region at an energy scale larger than superconducting gap in the $d^{2}I/dV^{2}$ spectra. Also, the peak/dip feature in $d^{2}I/dV^{2}$ spectra corresponds to the position where the dip-hump structure in conductance spectroscopy $dI/dV$ has the maximum magnitude of slope. The energy scales of the two boson modes arising from the double dip-hump structures discussed in the manuscript for the Bi-2212/1T-TaS$_{2}$ Junctions 1--3 could be obtained by numerically calculating the first order derivative of conductance spectroscopy $dI/dV$. A small amount of deviation is foreseeable since the conductance spectroscopy data are acquired in incremental steps of 1\thinspace mV. We argue that these deviations do not obscure the features of the double dip-hump structures. The $d^{2}I/dV^{2}$ spectra of Junctions\thinspace 1--3 at various temperatures up to 100\thinspace K are shown in Fig S.~\ref{Supp figure4}. Two dips in the negative bias region and two peaks in the positive bias region are marked by orange and purple arrows respectively, which are related to the maximum slopes of the two dip-hump structures in conductance spectroscopies at energy $\rm{E_{p}}$ and $\rm{E^{*}_{p}}$, as indicated by the dashed lines.

Bi-2212 is a \textit{d}--wave superconductor, the c-axis tunneling spectroscopies incorporate and average all possible gaps along momentum space. There is no consensus on the pairing mechanism of superconductivity in high-T$_{c}$ cuprates, and the consistency of the electron-boson interaction on nodal and anti-nodal regimes in momentum space is still unknown. Nonetheless, the good agreement of Bi-2212 intrinsic superconducting gap measured from Bi-2212/1T-TaS$_{2}$ and Bi-2212/graphite junctions with previous results on intrinsic Bi-2212 single crystal as discussed above and in the manuscript probably reveals the superconducting gap varies monotonically from the nodal to the anti-nodal regime along momentum space, which could clarify the rationality of the method for obtaining the boson mode energies from \textit{c}-axis conductance spectroscopies and $dI^{2}/dV^{2}$.

\section*{$I-V$ characteristics of Bi-2212/1T-TaS$_2$ (above $T_c$) and Ag/1T-TaS$_2$/Ag junctions}
\begin{figure}
\includegraphics[width=0.8\textwidth]{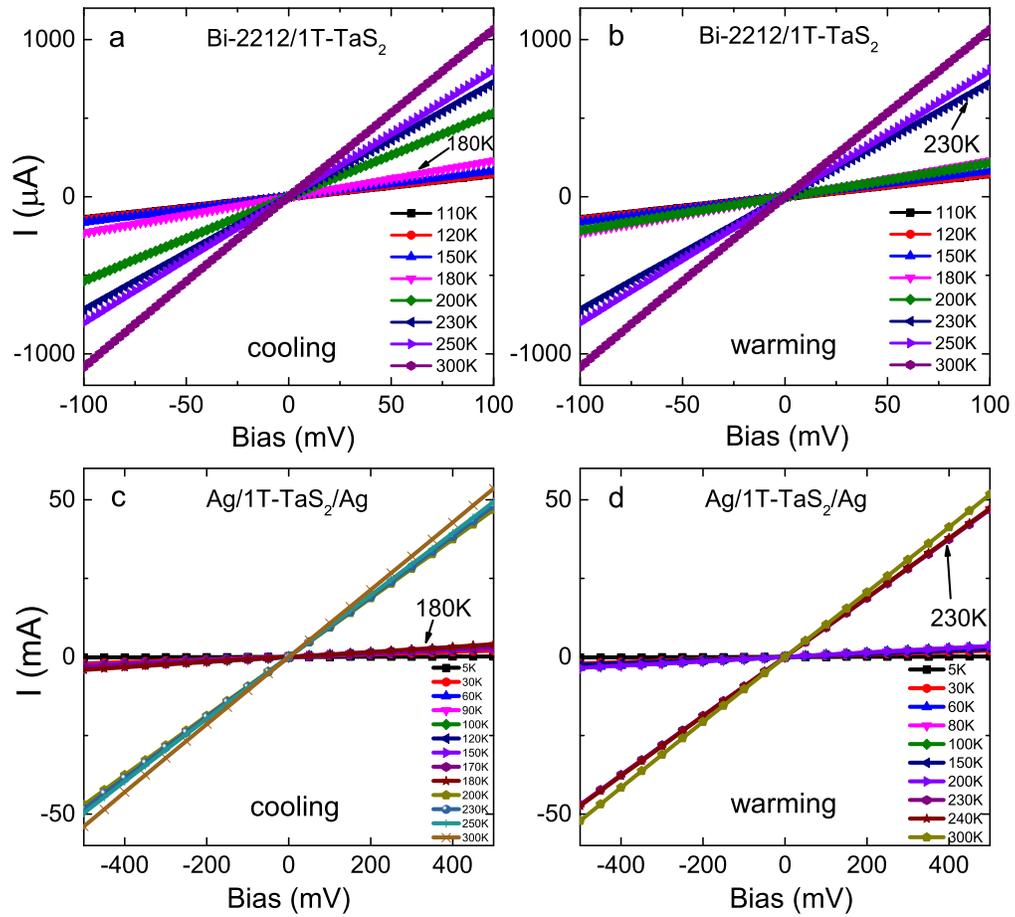}
\caption{$I-V$ characteristics of (a)--(b) Bi-2212/1T-TaS$_2$ Junction-1 (Z=0.25) at temperatures above Bi-2212's $T_c$ and (c)--(d) back-to-back structured Ag/1T-TaS$_2$/Ag junction for cooling and warming processes.}
\label{Supp figure5}
\end{figure}

DC $I-V$ characteristics of Bi-2212/1T-TaS$_2$ Junction-1 (Z=0.25) at temperatures above Bi-2212's $T_c$ and back-to-back structured Ag/1T-TaS$_2$/Ag junction under one cooling/warming cycle are shown in Fig.~S.~\ref{Supp figure5}. The Ag/1T-TaS$_2$/Ag junction are constructed by SPI Ag paste on 1T-TaS$_2$.

\clearpage

\end{document}